\documentclass[a4paper,11pt]{article}
\usepackage{jcappub} 
\usepackage{lineno}
\usepackage{mathtools}
\usepackage{macros}
\usepackage{comment}
\usepackage{amssymb,amsmath}
\usepackage{macros}
\usepackage{bm}
\usepackage{orcidlink}
\usepackage{subfig}
\usepackage{color}
\usepackage{amsthm}
\usepackage[update,prepend]{epstopdf}
\usepackage{hyperref}
\usepackage{float}
\usepackage{tabularx}
\usepackage{footnote}
\usepackage[normalem]{ulem}
\usepackage{soul}
\usepackage{cancel} 
\newcommand{\apjl}[1]{Astrophys. J. Lett.}

\begin{document}
\title{Lattice simulations of axion-U(1) inflation: gravitational waves, magnetic fields, and scalar statistics}
\date{\today}
\author[a]{Ramkishor~Sharma\orcidlink{0000-0002-2549-6861},}
\author[b,c,d,e]{Axel~Brandenburg\orcidlink{0000-0002-7304-021X},}
\author[f,g]{Kandaswamy~Subramanian\orcidlink{0000-0002-4210-3513},}
\author[a]{and~Alexander~Vikman\orcidlink{0000-0003-3957-2068}}
\affiliation[a]{CEICO, FZU-Institute of Physics of the Czech Academy of Sciences,
Na Slovance 1999/2, 182 00 Prague 8, Czech Republic}
\affiliation[b]{Nordita, KTH Royal Institute of Technology and Stockholm University, Hannes Alfv\'ens v\"ag 12, 10691 Stockholm, Sweden} 
\affiliation[c]{The Oskar Klein Centre for Cosmoparticle Physics, Department of Physics,\\ Stockholm University, AlbaNova, 10691 Stockholm, Sweden}
\affiliation[d]{McWilliams Center for Cosmology \& Department of Physics, Carnegie Mellon University, Pittsburgh, PA 15213, USA}
\affiliation[e]{School of Natural Sciences and Medicine, Ilia State University, 3-5 Cholokashvili Avenue, 0194 Tbilisi, Georgia}
\affiliation[f]{IUCAA, Post Bag 4, Ganeshkhind, Pune 411007, India}
\affiliation[g]{Department of Physics, Ashoka University, Rajiv Gandhi Education City, Rai, Sonipat 131029, Haryana, India}

\emailAdd{sharma@fzu.cz}
\emailAdd{brandenb@nordita.org}
\emailAdd{kandu@iucaa.in}
\emailAdd{vikman@fzu.cz}

\abstract
{We numerically study axion-U(1) inflation, focusing on the regime where
the coupling between axions and gauge fields results in significant
backreaction from the amplified gauge fields during inflation.
These amplified gauge fields not only generate high-frequency
gravitational waves (GWs), but also enhance spatial inhomogeneities in the
axion field. GWs serve as key probe for constraining the coupling
strength between the axion and gauge fields.
We find that, when backreaction is important during inflation, the
constraints on the coupling strength due to GW overproduction are relaxed compared to previous studies, in which backreaction matters only after inflation. Moreover, our results suggest that the probability density function (PDF) of axion fluctuations tends toward a Gaussian distribution even in cases where gauge field backreaction is important only after inflation. This aligns with previous studies where the same effect was observed for cases with strong backreaction during inflation.
This finding can be crucial for future studies of primordial black hole (PBH) formation, which can further constrain the coupling strength.
We also calculate the spectrum of the produced magnetic fields in this model and find that their strength is compatible with the observed lower limits.
}
\maketitle

\section{Introduction}
The idea of inflation originated from attempts to solve the singularity problem \cite{STAROBINSKY198099} and to address the horizon, isotropy and flatness 
problems in the hot big bang along with the monopole problem of grand unification, see \cite{Guth1981, LINDE1982389, PhysRevLett.48.1220} and \cite{Kazanas1980,SATO1981311}.
Inflation not only ameliorates the latter set of issues, but most importantly provides a natural mechanism \cite{Mukhanov:1981xt,Chibisov:1982nx,Mukhanov:2013tua} for generating initial density perturbations  in the universe from mere vacuum fluctuations, see also \cite{Starobinsky:1982ee,Guth:1982ec,HAWKING1982295,Bardeen:1983qw}. These perturbations account for anisotropies in the cosmic microwave background radiation (CMB) recently measured by Planck collaboration \cite{Planck:2018nkj} with unprecedented precision, thereby confirming the predictions of this mechanism and the theory of inflation. Later on these perturbations collapse to form all galaxies and the large-scale structure in the universe. 
Following the initial proposal, numerous models have been suggested
to realize an inflationary era in the early universe (for details see
refs.~\cite{Baumann:2009ds,Sriramkumar:2009kg,Martin:2013tda,Planck:2018jri,Chowdhury:2019otk}).
Axion inflation models are particularly interesting because they provide a natural high energy physics ingredient for a long lasting inflation - an approximate shift symmetry of the potential
\cite{Freese1990}. After the first proposal, several models have been suggested in the context of axion inflation \cite{Adams:1992bn,McAllister:2008hb,Kaloper:2008fb,Kim:2004rp,Anber:2009ua,Barnaby:2010vf,Barnaby:2011qe,Adshead:2012kp,Notari:2016npn,Ferreira:2017lnd}; see \RRef{Pajer:2013fsa} for a review. One of these models \cite{Anber:2009ua} considers a coupling of the axion field with the U(1) gauge field and we focus on this model in this paper; see \RRef{Alam:2024fid} for the observational status of axion-U(1)inflation.
The axion inflation model, where the axion couples to a gauge field,
exhibits a rich phenomenology.
This includes the production of gravitational waves (GWs)
\cite{Sorbo:2011rz,Anber:2012du,Adshead:2013qp,Adshead:2018doq,Adshead:2019aac,Adshead:2019igv,Adshead:2019lbr,Bastero-Gil:2022fme},
primordial black holes (PBHs)
\cite{Linde2013,Bugaev:2013fya,Domcke:2017fix},
matter-antimatter asymmetry \cite{Domcke:2019mnd,Domcke:2020kcp}, and primordial magnetic fields
\cite{Garretson:1992vt,Anber:2006xt,Fujita:2015iga,Adshead:2016iae,Kamarpour:2018ckk,Sobol:2019xls,Gorbar:2021zlr,Gorbar:2021rlt}.
Axion inflation with SU(2) gauge fields and its phenomenology has also been extensively explored \cite{Adshead:2012kp,Maleknejad:2012fw,Adshead:2013nka,Dimastrogiovanni:2012ew,Maleknejad:2016qjz,Maleknejad:2016dci,Dimastrogiovanni:2016fuu,Domcke:2018rvv,Maleknejad:2018nxz,Lozanov:2018kpk,Dimastrogiovanni:2018xnn,Wolfson:2020fqz,Iarygina:2021bxq,Fujita:2022jkc,Iarygina:2023mtj,Fujita:2023axo,Dimastrogiovanni:2024xvc,Dimastrogiovanni:2024lzj,Badger:2024ekb}.
The rich phenomenology of axion inflation has prompted extensive
studies to understand the dynamics of these models.
In scenarios with a sufficiently strong coupling between the axion and
the gauge field, the backreaction of the produced gauge fields becomes
important either during inflation or post inflation depending on the
coupling strength.
This model including the backreaction of the gauge fields has
been studied semi-analytically through perturbative approaches
\cite{Cheng:2015oqa,Domcke:2020zez,Peloso:2022ovc} and the gradient expansion
formalism (GEF) \cite{Sobol:2020lec,Gorbar:2021rlt}
under the assumption of a homogeneous axion field.
More recently, efforts have been made to numerically
investigate the strong backreaction regime, where
the assumption of homogeneous axion field is relaxed
\cite{Caravano:2022epk,Figueroa:2023oxc,Caravano:2024xsb}.
Motivated by the importance of inhomogeneities of the axion field, as
demonstrated in the numerical studies, these inhomogeneities have also
been incorporated perturbatively into GEF \cite{Domcke:2023tnn}.

This rich phenomenology of axion inflation has been used to constrain
the coupling strength between the axion and gauge fields.
The coupling between the axion and the gauge field results in the
production of gauge fields, which in turn
generate high-frequency GWs, typically in the GHz range \cite{Adshead:2019igv, Adshead:2019lbr}.
These GWs contribute to the early universe's total energy budget as additional radiation degrees of freedom.
The bound on additional radiation degrees of freedom from CMB observations constrains the produced GWs, which further
limits the coupling between the axion and the gauge fields.
We reexamine the bounds on the coupling strength between the axion and
the gauge field, considering stronger couplings than those explored in
previous works \cite{Adshead:2019igv,Adshead:2019lbr}.
In \cite{Adshead:2019igv}, the authors considered scenarios where the
backreaction due to the produced gauge field became important in the
post-inflationary era but not during inflation.
For one of their models, the chaotic inflation model, they found that the
coupling strength should be below $(m_\mathrm{pl}/70)^{-1}$ for an inflaton
mass equal\footnote{Here and throughout the paper $m_\mathrm{pl}=1.22\times10^{19}\,\,\rm{GeV}$ represents the Planck mass.} to $1.2\times 10^{-6} m_\mathrm{pl}$.
In our study, we explore scenarios with larger coupling strengths,
where the gauge field energy density becomes large enough to affect the
duration of inflation. In these scenarios,
inflation ends when the gauge field energy density
becomes comparable to the inflaton's potential energy.
Earlier studies of such scenarios, presented in  refs.~\cite{Caravano:2022epk,Figueroa:2023oxc},
did not address the production of GWs that we investigate in this paper. Our findings indicate that in such scenarios, the bounds on the coupling
strength are relaxed compared to previous estimates \cite{Adshead:2019igv,Adshead:2019lbr}.

Additionally, the produced gauge field can also enhance spatial fluctuations of the axion field. If these spatial fluctuations exceed a threshold value, they can lead to the formation of PBHs\footnote{
For other models of the formation of PBHs during inflation see, e.g., refs.~\cite{Khlopov:2008qy,Sasaki:2018dmp,Escriva:2022duf,Mishra:2019pzq,Ragavendra:2020sop}.
} in the post-inflationary era. This phenomenon has been analyzed in detail considering the $\chi^2$-distribution of the spatial fluctuations of the axion field. Given that the backreaction of the gauge field on the axion field is quadratic in nature, and assuming the gauge field has a Gaussian distribution, the spatial fluctuations of the axion field are expected to follow a $\chi^2$-distribution. By using constraints on PBH abundances, a bound on the axion and gauge field coupling has been obtained \cite{Linde2013,Bugaev:2013fya}.
A recent numerical study of axion inflation \cite{Caravano:2022epk}
considers the case when the backreaction of the gauge field on the axion
evolution becomes significant during inflation.
It was found that the distribution of the spatial fluctuations of the
axion field starts to deviate from the $\chi^2$-distribution and begins
to resemble a Gaussian distribution.
In this paper, we study the probability distribution of spatial
fluctuations in the axion field over a range of coupling strengths.
We find that the distribution tends to a $\chi^2$-distribution during
the evolution.
However, it gradually transitions to a Gaussian distribution at a later
time, irrespective of whether the backreaction of the gauge fields
becomes important during or after inflation.

The paper is organized as follows. In \Sec{section1}, we discuss the axion-inflation model and state the governing evolution equations for this system. In \Sec{section2}, we describe the initial conditions of our numerical simulations, present the simulation results, and compare them with previous semi-analytical studies. The production of GWs considered in \Sec{section4}. The magnetic part of the produced gauge fields may explain the presence of magnetic fields in the intergalactic medium, which we discuss in \Sec{section3}. In \Sec{section5}, we describe the probability density function of the curvature perturbation. In \Sec{section6}, we present the conclusions of our study.

Throughout this paper, the indices $i$, $j$, $k$ represent
three-dimensional vector indices, and the background spacetime metric
is assumed to be the spatially-flat Friedmann-Lemaitre-Robertson-Walker (FLRW) metric.
Capital bold letters denote three-dimensional vector quantities.
We use signature convention $(-,+,+,+)$ and Planck units $\hbar=c=G=1$.
The Planck mass is defined as ${m_\mathrm{pl}}=1/\sqrt{G}$.

\section{Axion-U(1) inflation model}\label{section1}

The following action describes the dynamics of the axion-U(1) inflation model,
\begin{align}\label{eq:first_Action}
    S=\int d^4x \sqrt{-g} \Bigg[\frac{m_\mathrm{pl}^2}{16 \pi}R-\frac{1}{2}\partial_\mu\phi \partial^\mu\phi-V(\phi)-\frac{1}{4}F_{\mu \nu}F^{\mu\nu}-\frac{\alpha}{4f}\phi F_{\mu \nu}\tilde{F}^{\mu\nu}\Bigg].
\end{align}
Here, $\phi$ represents the axion field and $V(\phi)$ denotes its
potential, which in our analysis is taken as\footnote{Although chaotic
inflation has been ruled out by CMB observations \cite{Planck:2018jri}, we have chosen this
model for several reasons.
First, it simplifies our analysis.
Second, it allows us to compare our results directly with those of
previous studies.
Lastly, many other inflationary potentials can be approximated by a
quadratic potential near the end of inflation, and our study primarily
focuses on this late phase of the inflationary period.}
$V(\phi) = m^2\phi^2/2$ with $m =1.06 \times 10^{-6} m_{\rm pl}$,
$F_{\mu\nu}=\partial_{\mu}A_{\nu}-\partial_{\nu}A_{\mu}$ represents the gauge field strength tensor,\footnote{In this work, we identify the gauge fields with the Standard Model hypercharge sector.
The production of fermions is important in this context \cite{Domcke:2018eki,Gorbar:2021rlt}, but their numerical modeling touches upon many unsolved questions, which we have to leave for future work.
Alternatively, the scenario in this study could apply to gauge fields in a dark sector, but in that case, these dark sector gauge fields would not provide magnetogenesis as discussed in \Sec{section3}.}
while $\tilde{F}^{\mu\nu}$ is its Hodge dual, 
$\alpha/f$ denotes the coupling strength between the axion field and
the U(1) gauge fields, and $R$ is the Ricci scalar.
Using this, and denoting derivatives with respect to conformal time
$\eta$ by a prime, we get the following dynamical equations\footnote{In this paper, we neglect the role of metric perturbations.
The terms that couple metric perturbations to the axion field are suppressed by slow-roll parameters and can thus be ignored during inflation \cite{Leblond:2010yq,Caravano:2022yyv}.
However, the role of these terms may become important near the end of inflation, which we leave for future investigation;
see refs.~\cite{Adshead:2023mvt,Galanti:2024jhw,Caravano:2024xsb} for recent studies on this topic.} for the axion-gauge
field evolution \cite{Adshead:2016iae}
\begin{align}
    \phi''+2\hhh \phi'-\ddel \phi+a^2 \frac{d V}{d \phi}&=\frac{\alpha}{f}\,\frac{1}{a^2}\,\EE\cdot\BB\label{phi_eqn}\,,\\
    \AA''-\Del A_0'-\ddel \AA+\Del(\Del \cdot \AA)-\frac{\alpha}{f}\left(\phi'\BB+\Del \phi \times \EE\right)&=0\,,\label{Ai_eqn}
\end{align}
where\footnote{We follow here the standard sign convention between $\EE$
and $\AA'$ \cite{Anber:2009ua}, which differs from that of some other
papers \cite{Adshead:2016iae, Figueroa:2023oxc}.}
$\EE=-\AA'+\Del A_0,~\BB=\Del \times \AA$ and
$\hhh$ is the comoving Hubble parameter with $\hhh=a'/a$ for scale factor $a(\eta)$.
On top of the constraint $\Del \cdot \BB=0$, the action \eqref{eq:first_Action} modifies the Coulomb law to
\begin{align}
    \Del \cdot \EE +\frac{\alpha}{f} \BB\cdot\Del \phi &=0\,.\label{A0_eqn}
\end{align}
Finally, the gravitational background obeys the Friedmann equations
\begin{align}
    \hhh^2=\frac{8\pi}{3 m_\mathrm{pl}^2}a^2 \rho\,, \quad\quad\quad\hhh'=-\frac{4\pi a^2}{m_\mathrm{pl}^2}(\rho+p)+\hhh^2\,,\label{hequation}
\end{align}
 where the density $\rho$ and pressure $p$ are given by \cite{Adshead:2019lbr}
\begin{align}
    \rho&=\Big\langle\frac{1}{2}\frac{\phi'^2}{a^2}+\frac{1}{2}\frac{(\Del\phi)^2}{a^2}+V(\phi)+\frac{\EE^2+\BB^2}{2a^4}\Big\rangle\,,\\
       p&=\Big\langle\frac{1}{2}\frac{\phi'^2}{a^2}-\frac{1}{6}\frac{(\Del\phi)^2}{a^2}-V(\phi)+\frac{\EE^2+\BB^2}{6a^4}\Big\rangle\,,
\end{align}
and the $\langle..\rangle$ represents the spatial averaging over a cubic
domain of size $L^3$, where $L=2\pi/k_1$ is specified in terms of the
lowest wave number of the domain, $k_1$. We consider different values for $L$, depending on the value of $\alpha/f$.

Before turning to the numerical results, it is useful to discuss here
the possibility of gauge field amplification in the case of a homogeneous axion field.
By assuming a homogeneous axion field ($\Del \phi=0$) and using the Coulomb gauge ($\Del \cdot \AA=0$), equation~\eqref{A0_eqn} implies $A_0=0$ and equation~\eqref{Ai_eqn} reduces to,
\begin{align}
\AA''-\ddel \AA-\frac{\alpha}{f}\phi'\Del \times \AA&=0\,.
\end{align}
By choosing the circular polarization basis, the above equation simplifies to the following form in Fourier space,
\begin{align}\label{Aequation_woback}
\left(\partial_\eta^2+k^2\mp 2 \xi (\hhh \eta)\frac{k}{\eta}\right)A_k^{\pm}&=0\,,\qquad \text{where} \qquad \xi=-\frac{\alpha}{2f}\, \,\frac{\phi'}{\hhh}\,. 
\end{align}
The above equation indicates that, as $\phi$ rolls down its potential,
one of the modes of the gauge field amplifies depending on the velocity
of the axion field ($\phi'/\hhh$) and the coupling strength ($\alpha/f$).
This phenomenon has been extensively discussed in the literature \cite{Garretson:1992vt,Fujita:2015iga,Adshead:2016iae,Gorbar:2021rlt}. For large coupling strengths, the backreaction from the produced gauge fields becomes significant and must be considered, as it alters the dynamics of the system.
This backreaction provides an additional friction term for the axion field evolution, leading to an extended duration of inflation compared to cases where backreaction is neglected.

In this paper, we use the {\sc Pencil Code} \cite{JOSS} to solve these
equations on a lattice in the regime where the backreaction from the
produced gauge fields is significant; see \App{AppendixA} for details
about the implementation of these equations.

\section{Numerical simulations of axion-U(1) inflation}\label{section2}

\subsection{Initial conditions}
\label{InitialConditions}

In this section, we describe the initial conditions for the axion and
gauge fields used in our simulations.
The initial conditions are set well within the inflationary era.
To achieve accelerated expansion of the universe, the pressure $p$
must satisfy $p<-\rho/3$ which, in the absence of gauge fields, implies $\phi'^2<a^2 V(\phi)$.
Using this condition in equations~\eqref{hequation}, we get
\begin{align}
    \hhh^2\approx\frac{8\pi}{3 m_\mathrm{pl}^2}a^2 V(\phi) \quad \text{and} \quad \epsilon_H\equiv-\frac{\dot{H}}{H^2}&\approx \frac{3}{2}\,\frac{\phi'^2}{a^2 V(\phi)}\,,
\end{align}
where $H=\dot a/a=\hhh/a$ is the usual Hubble parameter, dot denotes derivative with respect to cosmological time $dt=a d\eta$ and $\epsilon_H$ represents the first slow-roll parameter.
The above expressions imply
\begin{align}
    \phi'\approx a\sqrt{\frac{2\epsilon_H}{3}V(\phi)}\,.
\end{align}
Assuming $a=-1/(H \eta)$ during inflation, we get $\eta=-1/(a H)=-1/\hhh$,
and thus the initial values of $\phi$ and $\eta_i$ as
\begin{align}
    \phi'_{0}(\eta_i)=a_i\sqrt{\frac{2\epsilon_H}{3}V(\phi_{in})},\quad \eta_i=-\frac{1}{\hhh_i}\,.
\end{align}
Here $\hhh_i=\sqrt{[8\pi/(3 m_\mathrm{pl}^2)] V(\phi_0)}$ and $\phi_0$ is the initial
value of the homogeneous part of the axion field.
For the quadratic inflation model, $V(\phi)=m^2 \phi^2/2$, we have
$\epsilon_H=2/(8\pi \phi^2)$.
This implies,
 \begin{align}
    \phi'_{0}(\eta_i)=a_i\frac{m~m_{\rm pl}}{\sqrt{12 \pi}}\,.
\end{align}
In our simulations, we choose the initial condition for the axion field and its derivative such that
\begin{align}    \phi(\eta_i)&=\phi_0(\eta_i)+\delta\phi(\eta_i,\xx)\,,\\
\phi'(\eta_i)&=\phi_0'(\eta_i)+\delta\phi'(\eta_i,\xx)\,,
\end{align}
where $\delta\phi$ represents the fluctuations arising from the quantum mechanical nature of the axion field. We initialize these fluctuations as a
Gaussian random field with a power spectrum,
$P_{\delta\phi}(k)=\mathcal{H}^2/2k^3\left(k/\mathcal{H}\right)^2$
corresponding to the Bunch-Davies initial condition.
At the beginning of our simulation, the chosen box size ensures that
all wave numbers are within the Hubble horizon.
To avoid the effects of large ultraviolet modes, we cut off the spectrum
at a wave number $k_p$ to a decaying power law with a specific index.
Our results remain insensitive to the value of this cutoff wave number
$k_p$, as demonstrated in \App{kp_test}.
The initial spectra of $\delta \phi'$, $\AA$, and $\EE$ have been chosen correspondingly.
The spectra of the electric part, $P_E(k)$ and of the magnetic part,
$P_B(k)$ of the gauge field are defined via the comoving energy density,
\begin{equation}
\rho_g\equiv\frac{1}{2}\langle \EE^2+\BB^2 \rangle=\int \frac{dk}{k}\, [P_E(k)+ P_B(k)]\,.
\end{equation}

\begin{table}
\begin{center}
\begin{tabular}{ccccccc}
\hline
Run & $\alpha/f$ (in~${m_\mathrm{pl}^{-1}}$) & $\alpha/(\sqrt{8\pi}f)$ & $k_p/m$  &$k_1/m$ & grid size & $\Delta N_{e}$ \\
\hline
A & $35$ & $7$    & $47$ & $2$ & $512$ & $0.0$  \\ 
B & $50$ & $10$    & $47$ & $2$ & $512$ & $0.0$  \\ 
C & $60$ & $12$    & $47$ & $2$ & $512$ & $0.0$  \\ 
D & $75$ & $15$    & $47$ & $4$ & $1024$ & $1.7$  \\ 
D' & $75$ & $15$    & $94$ & $2$ & $512$ & $1.7$  \\ 
E & $90$ & $18$    & $47$ & $8$ & $1024$ & $4.0$  \\ 
\hline
\end{tabular}
\caption{Parameters for the runs discussed in the paper. 
Here $k_1$ represents the lowest wave number of the simulation domain
and $k_p$ is the cutoff wavenumber for the initial spectrum.
$\Delta N_e$ denotes the excess number of $e$-folds over the case without
backreaction due to the gauge fields.
}\label{table1}
\end{center}
\end{table}

\subsection{Dynamics of axion and gauge fields: inferred from simulations}\label{dynamics_gf}

\begin{figure}[h!]
\centering
 \includegraphics[width=1\textwidth]{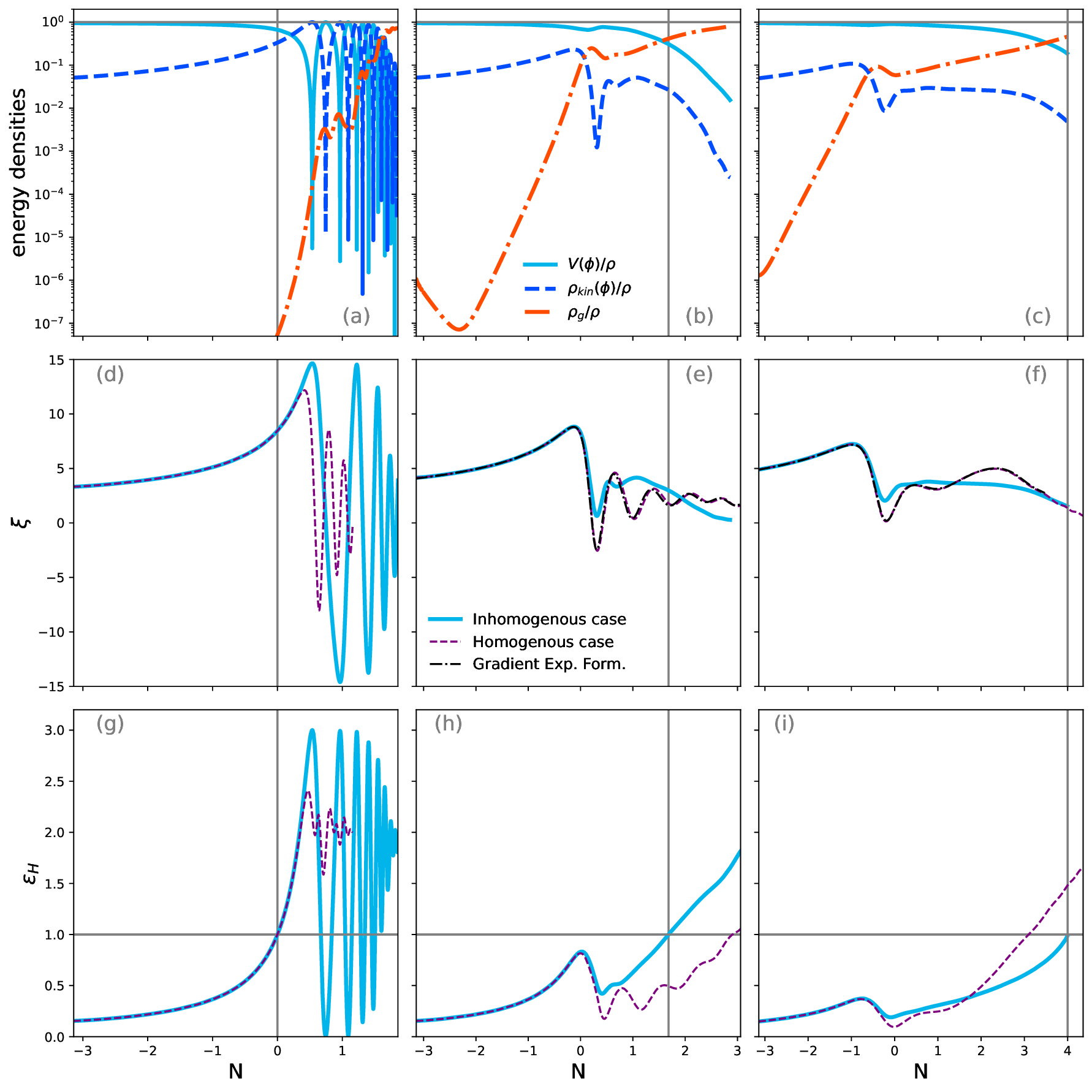}
\caption{The time evolution of different energy densities (top row),
$\xi$ (middle row), and the first slow-roll parameter, $\epsilon_H$
(bottom row), is shown for $\alpha/f = 60$ (left column), $\alpha/f = 75$ (middle
column), and $\alpha/f = 90$ (right column).
The vertical gray lines indicate the end of inflation for each case.
In the middle and bottom rows, the violet lines represent the time
evolution of $\xi$ and $\epsilon_H$ for the homogeneous $\phi$ case. In panels (e) and (f), the dot-dashed gray lines show the analytical
GEF results discussed in refs.~\cite{Sobol:2019xls,Sobol:2020lec}.}
\label{energy_densities}
\end{figure}

To study the dynamics of the axion-U(1) inflation model, we
perform a series of simulations with different coupling
strengths\footnote{It is to be noted that our coupling
strengths are in the units of the inverse of the Planck mass.
However, many previous studies use the inverse of the reduced Planck mass
for normalization.
To convert our coupling strengths to the reduced Planck mass
normalization, our values need to be divided by $\sqrt{8\pi}$.
\Tab{table1} provides the values of the coupling strengths in both units
for clarity.} $\alpha/f=30$, $50$, $65$, $75$, and $90$.
The parameters for these simulations are given in \Tab{table1}.
In \Fig{energy_densities}, we present the evolution of different energy
densities for $\alpha/f=60$, $75$, and $90$.
We show the evolution of the kinetic energy density, $\rho_\mathrm{kin}$
(dashed blue curves), potential energy density, $V(\phi)$ (solid cyan curves),
and gauge field energy density, $\rho_g$ (dot-dashed orange curves)
in panels (a), (b), and (c) of \Fig{energy_densities} for the coupling
strengths $\alpha/f = 60$, $75$, and $90$, respectively. Panels (d),
(e), and (f) illustrate the evolution of $\xi$, while panels (g), (h),
and (i) display the first slow-roll parameter, $\epsilon_H$, for the
corresponding cases. Inflation ends when $\epsilon_H = 1$. The vertical
gray lines indicate the epoch of the end of inflation in each case.
In this figure, $e$-folds $N$ is defined\footnote{As usual we define dimensionless $e$-folds $N$ through the relation $dN=Hdt$ with cosmological time $t$ and Hubble parameter $H$.} in such a way that $N=0$ would be the end of inflation
had one neglected backreaction, even in scenarios where backreaction is significant. We denote the corresponding excess number of $e$-folds as $\Delta N_{e}$. 

For $\alpha/f = 60$, inflation ends when the potential and kinetic
energies of the axion field become comparable and the backreaction of
the gauge field becomes important post-inflation.
The $\alpha/f = 30$ and $50$ cases have a similar behavior.
However, for $\alpha/f = 75$ and $\alpha/f =90$, the amplification of the gauge
fields is significant to backreact on the evolution of the axion field
during inflation itself.
When the backreaction of gauge fields on the axion field dynamics becomes
significant during inflation, it reduces the axion field velocity,
thereby extending the duration of inflation.
The duration of inflation extends by $\Delta N_{e}=1.7$ $e$-folds for $\alpha/f = 75$
and $\Delta N_{e}=4.0$ $e$-folds for $\alpha/f = 90$ compared to cases with smaller
coupling strengths.
Our results are similar to those obtained in \RRef{Figueroa:2023oxc},
except that here the extended duration of inflation $\Delta N_{e}$ is slightly shorter.

In scenarios where the backreaction of gauge fields is negligible
during inflation, it ends when the potential and kinetic energy become
comparable.
However, when gauge field backreaction is significant during inflation,
it ends when the gauge field energy density is comparable to the axion
potential energy. This conclusion is based on the numerical results of the simulations shown in \Fig{energy_densities}.
Similar findings have also been reported in the simulations of \RRef{Figueroa:2023oxc}.

\subsection{Comparison with previous work in the homogeneous $\phi$ case}
\label{ComparisonWithPreviousWork}

As discussed in the previous section, the backreaction of the gauge
fields on the evolution of the axion field becomes significant during
inflation for the coupling strengths; $\alpha/f=75$ or $90$.
Previous semi-analytical studies have studied the evolution of this by
ignoring spatial inhomogeneities \cite{Domcke:2020zez,Domcke:2023tnn,
Sobol:2020lec,Gorbar:2021rlt}, and more recently, lattice simulations have
been used to study this model \cite{Caravano:2022epk,Figueroa:2023oxc}.
In \RRef{Domcke:2020zez}, backreaction is incorporated
perturbatively, while the authors of \RRef{Sobol:2020lec} use
the gradient expansion formalism.
In this section, we compare our results with these semi-analytical findings.
\begin{figure}[h!]
\centering
 \includegraphics[width=1\textwidth]{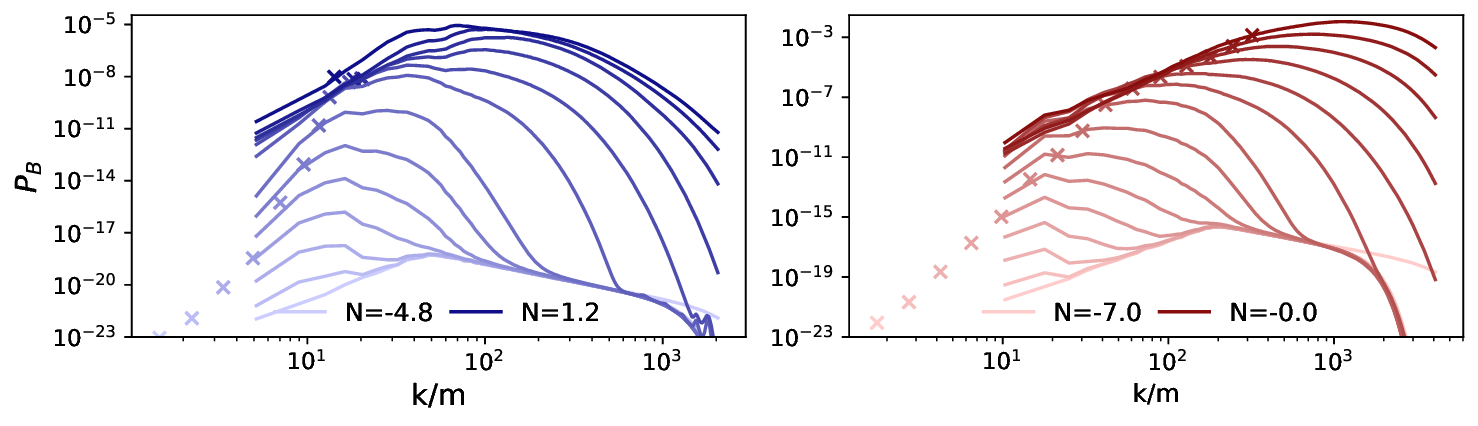}
 \includegraphics[width=1\textwidth]{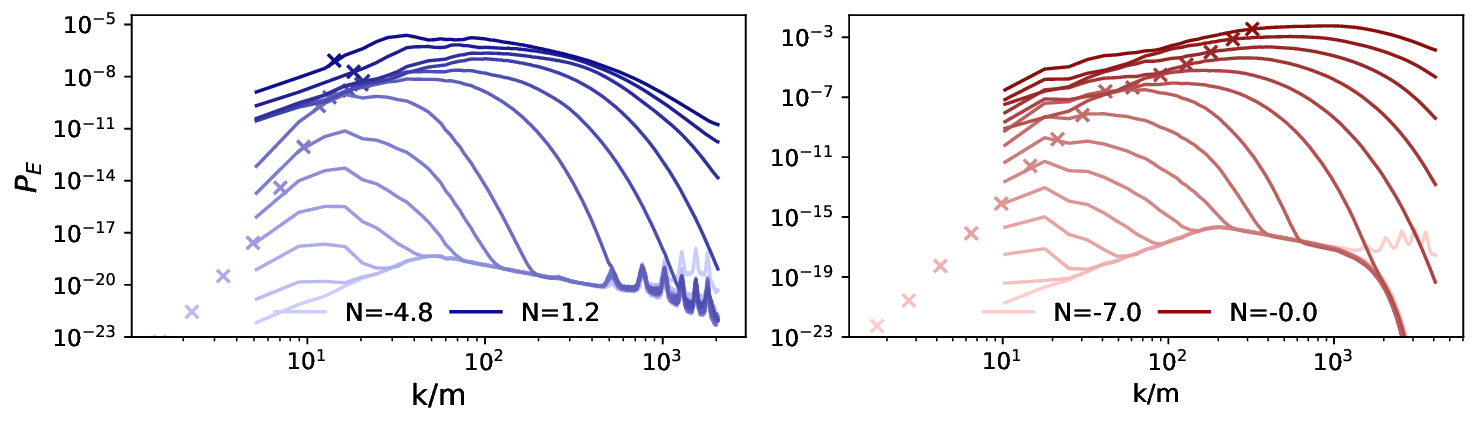}
\caption{
The spectra of the magnetic and electric components of the gauge field at
different times are shown for $\alpha/f = 75$ (left, blue curves, run D) and
$\alpha/f = 90$ (right, red curves, run E). Here, the cross (`x') represents the wavenumber corresponding to the Hubble horizon at the time of the spectrum.}
\label{magnetic_spectra}
\end{figure}

To compare our results with the previous semi-analytical studies, we
neglect the axion field inhomogeneity in the evolution of the axion and gauge
fields, i.e., equations~(\ref{phieq}), (\ref{efeq1}), and (\ref{gammaeq}),
and the contribution of the corresponding term in the total energy
budget that governs the evolution of the background FLRW geometry,
i.e., \Eq{hubbleeq}.
We compare our results with previous semi-analytical works using
a similar methodology to that in \RRef{Figueroa:2023oxc}.
In panels (d), (e), and (f) of \Fig{energy_densities}, the cyan curves
show the evolution of $\xi$, which governs the growth of the gauge fields.
The purple curves in these panels represent the evolution of $\xi$
when $\phi$ is assumed to be homogeneous, thus ignoring terms involving
the spatial derivatives of $\phi$ in the evolution equations.
The gray curve in panel (e) shows the predicted evolution of $\xi$
from the semi-analytical GEF model used to study the axion-U(1) system
described in refs.~\cite{Sobol:2020lec,Gorbar:2021rlt}.
From \Fig{energy_densities}, we conclude that our simulation results
align well with those of the GEF model.
To have a good match with the GEF results, we note that the initial
conditions for the simulation must be chosen so that the axion-U(1)
system reaches the slow-roll regime a few $e$-folds before the gauge field
backreaction becomes significant. This issue is discussed in detail in
\App{app_comp_with_gef}.

For the coupling strengths $\alpha/f=75$ and $90$, the spatial
fluctuations of gauge fields become prominent and suppress the oscillatory
features observed in the homogeneous case (purple curves in panels (e)
and (f)).
This attenuation has also been demonstrated in \cite{Figueroa:2023oxc}.

In \Fig{magnetic_spectra}, we show the spectra of the magnetic and
electric components of the gauge fields for run D (blue curves) and
run~E (red curves).
For run D, the spectra are shown from the initial time $N_i = -4.8$
to the final time $N_f = 1.2$, with intervals of $\Delta N = 0.5$.
For run E, the spectra span from $N_i = -7$ to $N_f = 0$,
which is also the time span shown in \RRef{Figueroa:2023oxc}.
We show the spectra of scalar field fluctuations at different times for
these runs in \App{scalar_spec}.

As seen in \Fig{magnetic_spectra}, both magnetic and electric energy
spectra initially peak at a scale close to the minimum wave number of
the simulation box.
As time progresses, the spectral peaks shift towards higher wave numbers,
eventually reaching close to the maximum box wave number for run E.
To extend the simulation to later times, a larger simulation box would
be required. For the simulation shown in \Fig{magnetic_spectra}, we note that
there is a late-time growth in the electric and magnetic energy spectra for superhorizon wave numbers.
A similar growth has also been seen in other simulations \cite{Figueroa:2023oxc,Figueroa:2024rkr}.

\section{Gravitational waves}\label{section4}

In this section, we discuss the production of GWs during axion-U(1)
inflation.
The interaction between the axion and the gauge field results in the
generation of GWs, as the produced gauge field has a nonzero transverse
traceless component in its energy-momentum tensor.
Additionally, scalar field spatial fluctuations produced by the gauge
field also contribute to the generation of GWs.
GWs are defined as the transverse traceless part of the tensor
perturbations ($h_{ij}^\mathrm{TT}$) of the metric.
Including these tensor perturbations, the metric is given by
\begin{align}
    ds^2=a^2\left(-d\eta^2+(\delta_{ij}+h_{ij}^\mathrm{TT})dx^i dx^j\right)
\end{align}
The evolution of these tensor perturbations is governed by the Einstein equation. By linearizing the Einstein equation, we obtain
\begin{align}
    \left(\frac{\pp}{\partial \eta^2}+k^2-\frac{a''}{a}\right)\tilde{h}_{ij}=\frac{16 \pi}{m_\mathrm{pl}^2 }\,\frac{\tilde{T}_{ij}}{a},
\end{align}
where, $\tilde{h}_{ij}=a h_{ij}^\mathrm{TT}$,
$\tilde{T}_{ij}=a^2T_{ij}^\mathrm{TT}$ and
$T_{ij}^\mathrm{TT}$ denotes the transverse traceless part of the
energy-momentum tensor ($T_{ij}$) of the source that led to the production
of GWs.
For the axion-U(1) model, the total $T_{ij}$ is given by
\begin{align}\label{eqn:Tij}
T_{ij}=-B_i B_j- E_i E_j + a^2 \partial_i \phi \partial_j \phi+...
\end{align}
Here, the ellipsis represents the irrelevant terms proportional to $g_{ij}$. The energy density of the produced GWs is given by
\begin{align}
    \rho_\mathrm{GW}\equiv\frac{1}{32 \pi a^2} \langle h_{ij}^{'\mathrm{TT}} h_{ij}^{'\mathrm{TT}}\rangle=\frac{1}{32 \pi a^4} \big(\langle \tilde{h}'_{ij} \tilde{h}'_{ij}\rangle-2 \hhh \langle \tilde{h}_{ij} \tilde{h}'_{ij}\rangle+\hhh^2\langle \tilde{h}_{ij} \tilde{h}_{ij}\rangle\big)\,.
\end{align}
The density fraction of GWs at the present epoch is given by
\begin{align}
\Omega_\mathrm{GW}h^2&=\frac{\rho_\mathrm{GW}}{\rho_c}\Big|_0 h^2=\frac{\rho_\mathrm{GW}}{\rho_r}\Big|_0 \Omega_r h^2=\frac{\rho_\mathrm{GW}|_e a_e^4}{\rho_r|_0 a_0^4} \Omega_r h^2.
\end{align}
In the above expression, we use $\rho_r|_0=\Omega_r \rho_c|_0$ and $\rho_{GW}|_0=\rho_{GW}|_e (a_e/a_0)^4$. Here, $\Omega_r$ represents the radiation density fraction at the
present epoch, $\rho_{GW}$ denotes the GW energy density at the epoch `e'\footnote{Here, `e' refers to an epoch when the Universe is radiation-dominated and after this epoch, there is no further production of GWs.} and $a_e$ and $a_0$ are the scale factors at epoch `e' and the present epoch, respectively.. By considering the adiabatic evolution of the Universe, i.e., $g_s^{1/3} a T = \mbox{const}$, we obtain
\begin{align}
\Omega_\mathrm{GW}h^2&=\frac{\rho_\mathrm{GW}|_e g_{0,s}^{4/3} T_0^4}{\rho_r|_0 g_{e,s}^{4/3} T_e^4} \Omega_r h^2.
\end{align}
Here, $T_0$ and $T_e$ represent the temperature of the Universe at the present and at the epoch `e', respectively. Furthermore, $g_{e,s}$ and $g_{0,s}$ denote the effective
degrees of freedom in the entropy density at the epoch `e' and the present
epoch, respectively. Further, by using the radiation energy density at the present epoch, $\rho_r|_0=g_0 \pi^2 T_0^4/30$, and at the epoch `e', $\rho_r|_e=g_e \pi^2 T_e^4/30$, we obtain
\begin{align}\label{omegagw_expr}
\Omega_\mathrm{GW}h^2&=\left(\frac{g_{0,s}}{g_{e,s}}\right)^{4/3}\frac{g_e}{g_0}\frac{\rho_\mathrm{GW}}{\rho_r}\Big|_e\Omega_r h^2\, \nonumber\\
 &=0.4 \times \left(\frac{g_{0,s}}{3.94}\right)^{4/3}\frac{g_e}{106.75}
 \left(\frac{g_0}{3.36}\right)^{-1} \left(\frac{g_{e,s}}{106.75}\right)^{-4/3}
 \frac{\rho_\mathrm{GW}}{\rho_r}\Big|_e\Omega_r h^2\, ,
\end{align}
where $g_e$ and $g_0$ represent the radiation degrees of freedom in energy density at the `e' and present epochs, respectively. The above expression represent $\Omega_\mathrm{GW}h^2$ in terms of the ratio of the mean GW energy density and the total energy density at the epoch `e', $\rho_\mathrm{GW}/\rho_r|_e$.

\begin{figure}[h!]
\centering \includegraphics[width=1\textwidth]{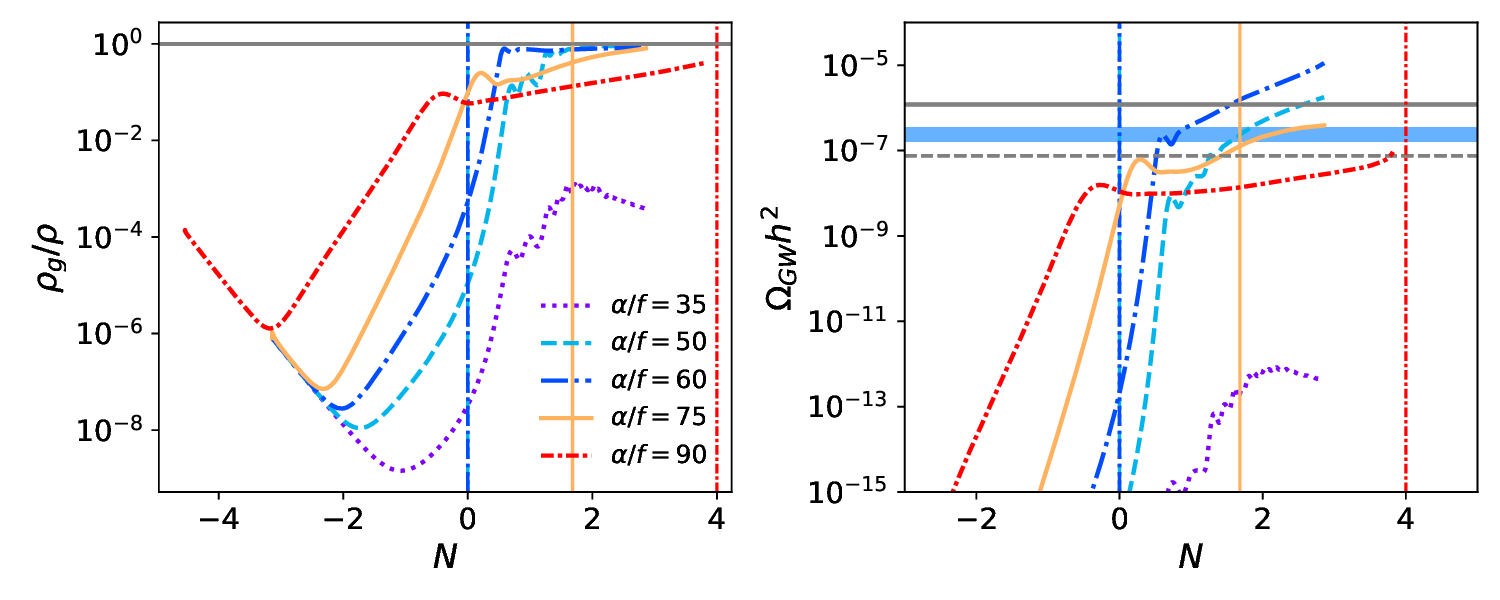}
\caption{
The time evolution of the gauge field energy density (left panel), normalized by the total energy density, and the density fraction of GWs (right panel) is shown for different values of $\alpha/f$. The vertical lines indicate the end of inflation for each case. In the right panel, the horizontal solid gray line represents the bound on $\Omega_\mathrm{GW}h^2$ from Planck observations of the CMB, while the dashed gray line denotes the combined sensitivity of the future mission COrE and Euclid \cite{Pagano:2015hma}. The blue band represents the CMB's S4 1$\sigma$ and 2$\sigma$ projections \cite{Abazajian:2019eic}.}
\label{gw_plot}
\end{figure}

Further in this section, we study the production of GWs in the axion-gauge field scenario, considering both weak and strong coupling cases.
The characteristic current frequency of these produced GWs is in GHz range and is determined by the horizon scale at the end of inflation, see equation \eqref{eq:Horizon_Inf}. 
In \Fig{gw_plot}, we show the time evolution of the gauge field energy density normalized by the total energy density in the left panel and the density fraction of the GWs in the right panel.
The curves correspond to different coupling strengths; $\alpha/f = 30$ (dotted violet), 50 (dashed blue), 60 (dot-dashed blue), 75 (solid orange), and 90 (dot-dashed red).
The vertical lines, in corresponding colors, represent the end of inflation for each case.
The horizontal solid gray line indicates the bound on the GW energy density from the $\Delta N_\mathrm{eff}$ constraints from the CMB.
The dashed gray curve represents the sensitivity of the future CMB-S4 mission. In the right panel of \Fig{gw_plot}, we use the ratio of the GW energy density to the total energy density obtained from our simulations as $\rho_\mathrm{GW}/\rho_r|_e$ in \Eq{omegagw_expr} to determine the time-dependent value of $\Omega_{\text{GW}}$ shown on the $y$ axes.
Our key findings regarding the constraints on the coupling strength between axion and gauge fields from the CMB bound on $\Delta N_\mathrm{eff}$ can be summarized as follows.

\begin{itemize}
\item As shown in \Fig{energy_densities}, for $\alpha/f = 75$ and $\alpha/f = 90$, once the energy density of the gauge field becomes comparable to the kinetic energy of the axion field, the velocity of the axion decreases. This reduction in axion velocity leads to a slower growth rate of the gauge field compared to the pre-backreaction stage. Consequently, the production rate of GWs also slows down, as shown in \Fig{gw_plot}. The amplitude of the generated GWs remains below the bound obtained using Planck data for these cases up until the end of inflation.

\item Previous studies \cite{Adshead:2019igv,Adshead:2019lbr} indicated
that the coupling strength above $\alpha/f = 70$ leads to an
overproduction of GWs, violating the $\Delta N_\mathrm{eff}$ bound from the
Planck data of CMB.
This bound is for the case of chaotic potential, which is also considered in this work.
However, we find that this bound is relaxed in the strong coupling case
when gauge field backreaction becomes significant during inflation.
The specific details of this bound depend on the duration of GW production
after the end of inflation.
As evident from \Fig{gw_plot}, the produced GWs for $\alpha/f=75$
may violate the bound from the Planck data on the GW amplitude, but
this occurs 2--3 $e$-folds after the end of inflation.
In our analysis, we have not considered the interaction of the gauge field
with other standard model particles.
For large coupling strengths, this interaction becomes important and
may lead to instantaneous reheating, as discussed in section 4.2 of
\RRef{Adshead:2016iae}.
Considering these effects would lead to the decay of the gauge field
into other standard model species, further affecting GW production.
If the gauge field energy density decays significantly within 2--3
$e$-folds after inflation, the produced GW amplitude will be smaller and
may well be within the bound obtained from the Planck data (shown by the
solid horizontal gray line in \Fig{gw_plot}).

In addition, we observe that, for the cases $\alpha/f=75$ and $\alpha/f=90$, the growth rate of the gauge field energy density decreases at the time when the backreaction of the gauge field becomes significant.
This is shown in the left panel of \Fig{gw_plot} as solid yellow and dash-dotted red curves.
This allows more time for the production of charged particles compared
to cases where backreaction becomes important only after inflation, such as $\alpha/f=50$ (dashed cyan curves) and $60$ (dash-dotted blue curves) shown in \Fig{gw_plot}.
From this discussion, we expect that the bound \cite{Adshead:2019igv,Adshead:2019lbr} on $\alpha/f$ coming from Planck datasets will be relaxed.  
Indeed, assessing the status of the bound requires the inclusion of the charge current (which depends on charge particle content and high energy physics at energy scales $10^{14}$~GeV) which can further only suppress the production of the gauge fields and consequently GWs.
A detailed study including the effects of charged particles is beyond the scope of this paper. We will report that in a future publication.

\begin{figure}[t]
\centering \includegraphics[width=1\textwidth]{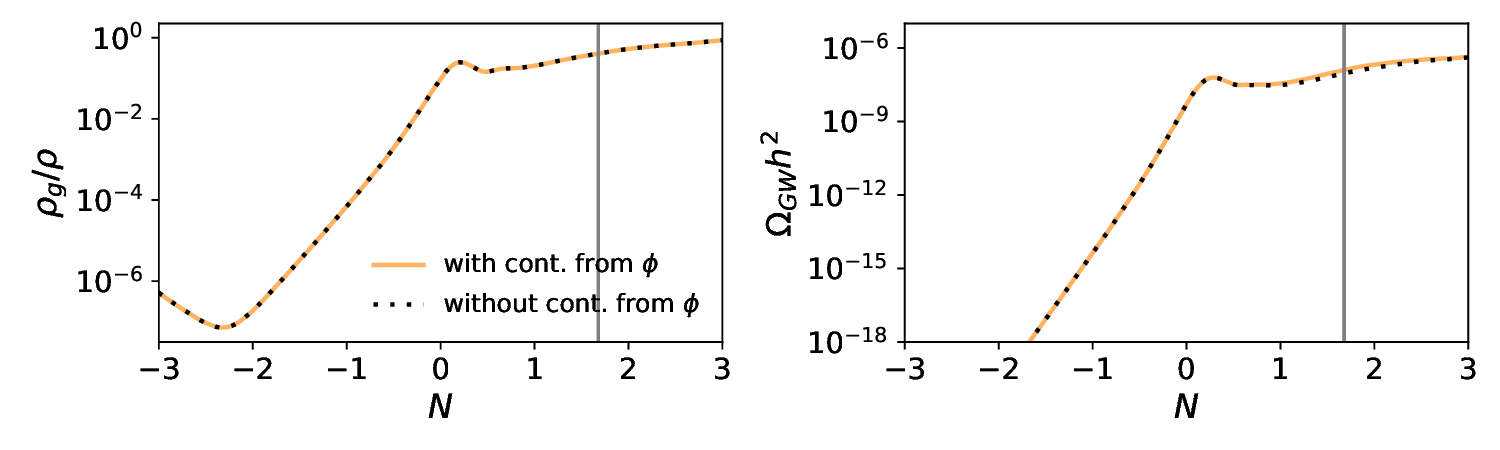}
\caption{Time evolution of $\rho_g/\rho$ and $\Omega_{GW}h^2$ for
$\alpha/f=75$ for run~D (solid yellow curves) and for a run (dotted black
curves) analogous to run~D in which the contribution of the $\phi$ term
in the total energy-momentum tensor, given in equation\eqref{eqn:Tij},
is neglected in the calculation of the GW energy density.}
\label{gw_plot1}
\end{figure}

\item Both axion and gauge fields contribute to GW production.
To compare their contributions, we conduct a simulation analogous to run D, but with the axion field contribution neglected.
We demonstrate this in \Fig{gw_plot1}.
The left panel shows the time evolution of $\rho_g/\rho$, while the right panel is for $\Omega_\mathrm{GW} h^2$.
The solid yellow curves correspond to the evolution for run D, whereas
the dotted black curves represent a similar run where the contribution of
the $\phi$ field was neglected in the estimation of the GW energy density.
The vertical gray line marks the end of inflation for both runs.
From this figure, we conclude that the contribution of the axion field
to the GW energy can be neglected for most of the duration.

\end{itemize}

\section{Magnetogenesis}\label{section3}
There is indirect evidence of magnetic fields in the intergalactic medium,
inferred from the non-detection of secondary GeV photons from blazars \cite{Neronov:1900zz, Taylor:2011bn, Finke:2013tyq, Finke:2015ona}.
The presence of such intergalactic magnetic fields motivates the study
of their generation in the early universe, potentially during inflation
\cite{Turner:1987bw, Ratra:1991bn, Demozzi:2009fu, Martin:2007ue,Ferreira:2013sqa,
Sharma:2017eps, Kushwaha:2020nfa, Kushwaha:2022bwy, Maity:2021qps, Tripathy:2021sfb,Brandenburg:2024awd}
or a first-order phase transition \cite{Hogan1983, Quashnock1989,
Vachaspati:1991nm, Baym:1995fk, Sigl:1996dm}, for reviews see
\RRef{Durrer:2013pga, Subramanian:2015lua, Vachaspati:2020blt}.
In this section, we study the generation of
magnetic fields within the axion-U(1) inflation scenario and calculate their present-day strength for different coupling strengths, $\alpha/f$.

\subsection{Gauge-field production}\label{GaugeFieldProduction}

As discussed in \Sec{dynamics_gf}, the gauge fields are amplified due
to the coupling between the axion and gauge fields.
The produced gauge fields consist of two components; the magnetic part
(hypermagnetic fields) and the electric part (hyperelectric fields).
The energies in the electric component of the gauge fields will be
converted into kinetic and magnetic energies \cite{Bran+Prot23} within
one Hubble time after reheating due to the high conductivity of the
universe post-reheating.
By contrast, the hypermagnetic fields remain frozen because of
the medium's high conductivity.
However, this is not entirely accurate as hypermagnetic fields also decay
as a power law due to their interaction with the charged plasma via the
Lorentz force \cite{BEO96, Mattias2001, Banerjee:2004df, Brandenburg:2016odr,
Subramanian:2015lua}, which we discuss later in this section.
At the electroweak phase transition, these hypermagnetic fields will
convert to standard magnetic fields, with their strength related to the
hypermagnetic fields by the Weinberg angle, which is $\mathcal{O}(1)$.
If strong enough, these magnetic fields could explain the presence of
intergalactic magnetic fields in the present-day universe.
We discuss this later in this section.
The possibility of magnetogenesis in axion inflation has been studied
in \RRef{Adshead:2016iae} for coupling strengths $\alpha/f = 60$ and below.
In this study, we also calculate the magnetic field strengths for larger
coupling strengths; $\alpha/f=75$ and $90$ \cite{Figueroa:2023oxc}
in addition to the cases considered in \cite{{Adshead:2016iae}}.

First, we study the evolution of the electric and magnetic components of the gauge fields during inflation in our simulations.
The root mean square value of the hypermagnetic field and its coherence length are given by
\cite{Durrer:2013pga, Subramanian:2015lua},
\begin{align}
B_\mathrm{rms}=\frac{1}{a^2}\sqrt{{\int d\log k \cdot P_{B}(k)}}\,\,, \quad\quad {\rm and}\quad\quad L_{c} =\frac{\int d\log k\cdot k^{-1}\cdot P_{B}}{\int d\log{k}\cdot P_{B}}\,\,.
\end{align}

Similar definitions are used for the electric part of the gauge fields.
Furthermore, assuming the universe evolves in a radiation-dominated
phase after epoch `e' with no further magnetic field generation, we can calculate the magnetic
field strength today using the following expression,
\begin{align}\label{brms_value}
B_\mathrm{rms}\big|_0&=B_{rms}|_e \left(\frac{a_e}{a_0}\right)^2=B_{rms}|_e \left(\frac{g_{0,s}^{2/3}T_{0}^2}{g_{e,s}^{2/3} T_e^2}\right)=\sqrt{\frac{{\int d\log{k}\cdot P_{B}}}{\rho_e}} \sqrt{\frac{\pi^2 g_e}{30}}\left(\frac{g_{0,s}}{g_{e,s}}\right)^{2/3} T_0^2.\nonumber\\
&=1.9\, {\rm \mu G}\,\sqrt{\frac{{\int d\log{k}\cdot P_{B}}}{\rho_e}} \left(\frac{g_e}{106.75}\right)\left(\frac{g_{0,s}}{3.94}\right)^{2/3}\left(\frac{106.75}{g_{e,s}}\right)^{2/3}\left(\frac{T_0}{2.73 {\rm K}}\right)^2
\end{align}
In the above expression, we assume the adiabatic evolution of the Universe, given by $g_s^{1/3} a T =const$, and use $\rho_e=g_e \pi^2 T_e^4/3$. Here, $\rho_e$ represents the total energy density of the universe at epoch `e', and $T_0$ denotes the CMB temperature.

\begin{figure}[h!]
\centering
 \includegraphics[width=1\textwidth]{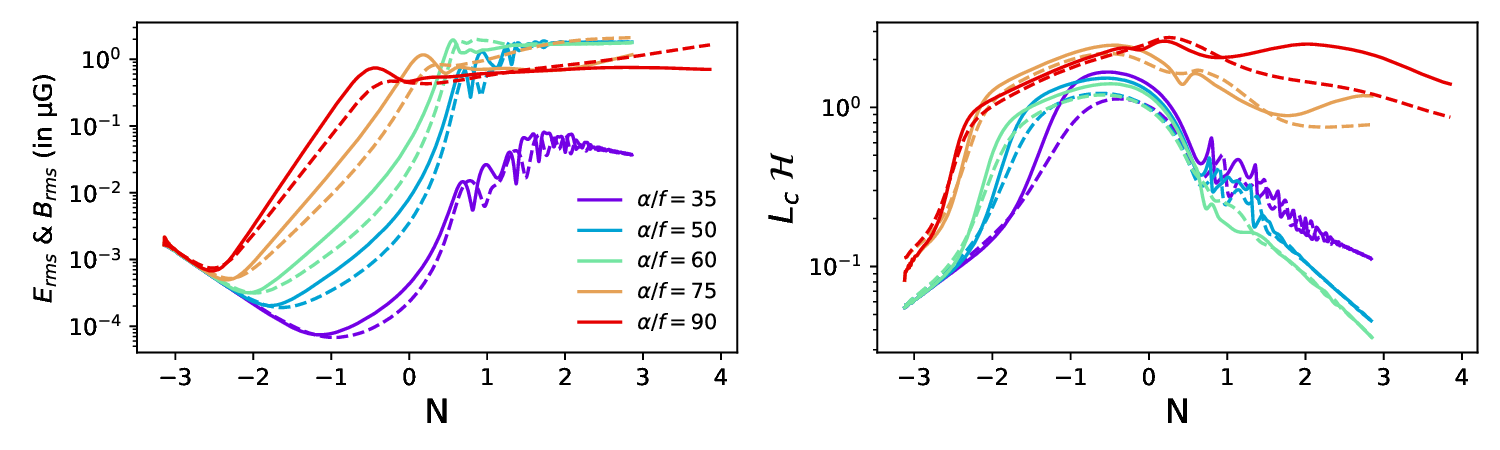}
\caption{The evolution of the electric
(solid lines)
and magnetic (dashed lines) components of the gauge field (left panel) and their coherence lengths (right panel) is shown for different values of $\alpha/f$. The coherence lengths are normalized by the length scale corresponding to the comoving Hubble rate ($\mathcal{H} = a'/a$) as a function of time.}
\label{magnetic_field_and_coh_len}
\end{figure}

In \Fig{magnetic_field_and_coh_len}, we demonstrate the evolution of
the electric component with solid lines and the magnetic component with
dashed lines for coupling strengths $\alpha/f = 30$ (green curve),
$50$ (red curve), $60$ (orange curve), $75$ (blue curve), and $90$
(yellow curve).
The values of $E_\mathrm{rms}$ and $B_\mathrm{rms}$ on the $y$-axis in the left panel are determined using the ratio $\sqrt{\int d\log{k}\cdot P_{B}/\rho_e}$ obtained from the simulation in \Eq{brms_value}.
In the right panel, we illustrate the evolution of the coherence length
for each coupling strength.
The coherence length scale is normalized by the length scale corresponding
to the comoving Hubble rate, $\mathcal{H}$, as a function of time.
The length scale corresponding to the Hubble horizon at the end of
inflation is given by
\begin{align}
\label{eq:Horizon_Inf}
L_H=\frac{a_0}{a_e} \frac{1}{H}=1.97 \times 10^{-17} {\rm pc} \left(\frac{10^{-6}m_{\rm pl}}{H}\right)^{1/2}.
\end{align}
In the above expression, we assume that the universe evolves adiabatically after inflation,
\begin{align}\label{sf_ratio}
\frac{a_e}{a_0} = \left(\frac{g_{0,s}}{g_{e,s}}\right)^{1/3}\frac{T_0}{T_e} = 2.66 \times 10^{-29} \left(\frac{g_{0,s}}{3.94} \frac{106.75}{g_{e,s}}\right)^{1/3}\left(\frac{106.75}{g_e}\right)^{1/4} \frac{T_0}{2.73 \text{K}} \left(\frac{ 10^{-6} m_{\rm pl}}{H}\right)^{1/2},
\end{align}
where $g_{e,s}$ and $g_{0,s}$ denote the effective
degrees of freedom in the entropy density at the end of inflation and the present
epoch, respectively. We estimate the reheating temperature, $T_r$, by assuming instantaneous reheating using
$3H^2 m_\text{pl}^2 = 8\pi(\pi^2/30)g_eT_e^4$.

\subsection{Evolution of gauge fields post inflation}\label{NLP_section}

After the end of inflation, the universe is dominated by the gauge field
in the strong coupling regime of axion-U(1) system.
Due to the dominance of the gauge field in the total energy budget,
the effective equation of state resembles that of radiation dominance.
We assume that reheating in this model occurs through the interaction
of the gauge field with other Standard Model (SM) fields, as described in
\cite{Adshead:2015pva, Adshead:2016iae}.
Therefore, we can consider the universe as being radiation-dominated
just after the end of inflation.

As discussed earlier, the gauge fields have a blue spectrum at the end
of inflation and peaks at a scale smaller than the Hubble horizon size; see \Fig{magnetic_spectra}.
We assume that the magnetic part of the gauge fields remain unaffected \footnote{Thus, we effectively assume that only the electric part of the gauge fields energy gets transferred to the SM particles. This is a simplified scenario. In particular one can expect that an important role should be played by the charge currents. Moreover, depending on the reheating scenario magnetic energy may also be reduced. However, the exact mechanism of reheating and energy transfer lies outside of the scope of this work.} from the end of
inflation until the epoch when the universe's conductivity becomes much
larger than the Hubble parameter. 
Once this epoch is reached, the electric component of the gauge field
shortened out within a Hubble time scale, and the magnetic field becomes
frozen into the rest of the plasma.
After this epoch, the magnetic field interacts with the plasma via the
Lorentz force and transfers some of its energy to the plasma.
As a result, the magnetic field strength decreases.
This nonlinear processing of the magnetic field begins when the Alfv\'en
time becomes smaller than the Hubble time (see section~6 of \RRef{Subramanian:2015lua} for details),
which implies
that the corresponding scale factor is
$a_\textrm{nl}\equiv a_H(\mathcal{H}~L_c/V_A)$.
Here, $a_H$ corresponds to the scale factor at the horizon entry for
a particular scale, and $V_A=\sqrt{B^2/(\rho+p)}$ is the Alfv\'en
velocity.

\begin{table}
\begin{center}
\begin{tabular}{c|c|p{1.8cm}|p{1.8cm}|p{1.8cm}|p{1.8cm}|p{1.8cm}|p{1.8cm}}
\hline
Run & $\alpha/f$ &$B_\mathrm{rms}|_0$ (G)&$L_c^\mathrm{NL}|_0$ (pc)&$B_\mathrm{bound}$ (G)&$B^1_\mathrm{rms}|_0$ (G)&$L_c^\mathrm{1,NL}|_0$ (pc)&$B^1_\mathrm{bound}$ (G) \\
\hline
A & $35$ & $6.2 \times 10^{-17}$ & $2.0 \times 10^{-3}$ & $1.8 \times 10^{-13}$    & $1.1 \times 10^{-15}$ & $1.3 \times 10^{-2}$ & $7.0 \times 10^{-14}$\\ 
B & $50$ & $4.5 \times 10^{-16}$ & $1.2 \times 10^{-2}$ & $7.3 \times 10^{-14}$    & $1.3 \times 10^{-14}$ & $1.5 \times 10^{-1}$ & $2.1 \times 10^{-14}$\\ 
C & $60$ & $1.7 \times 10^{-15}$ & $4.1 \times 10^{-2}$ & $4.0 \times 10^{-14}$    & $2.7 \times 10^{-14}$ & $1.5 \times 10^{-1}$ & $2.1 \times 10^{-14}$\\ 
D & $75$ & $1.5 \times 10^{-14}$ & $2.5 \times 10^{-1}$ & $1.6 \times 10^{-14}$    & $2.6 \times 10^{-14}$ & $2.4 \times 10^{-1}$ & $1.7 \times 10^{-14}$\\ 
E & $90$ & $1.9 \times 10^{-14}$ & $2.5 \times 10^{-1}$ & $1.6 \times 10^{-14}$    & -- & -- & --\\ 
\hline
\end{tabular}
\caption{
$B_\mathrm{rms}|_0$ and $L_c^\mathrm{NL}|_0$ represent the present-day
magnetic field strength and its coherence length obtained for each run,
respectively. These $B_\mathrm{rms}|_0$ and $L_c^\mathrm{NL}|_0$ in columns 3 and 4 are calculated from the spectrum at the end of inflation. For the sake of comparison, in columns 6 and 7 we provide $B_\mathrm{rms}^{1}|_0$ and $L_c^{1,\mathrm{NL}}|_0$ calculated from the spectrum obtained one $e$-fold after the end of inflation, provided we have these data. $B_{\rm bound}$ denotes the lower bound on the magnetic
field strength based on blazar observations, where we assume
$B_{\rm bound} =1.8 \times 10^{-17} {\rm G} \left(L / 0.2 \text{Mpc}\right)^{1/2}$ \cite{MAGIC:2022piy}.
}\label{table2}
\end{center}
\end{table}

\begin{figure}[h!]
\centering
 \includegraphics[width=0.7\textwidth]{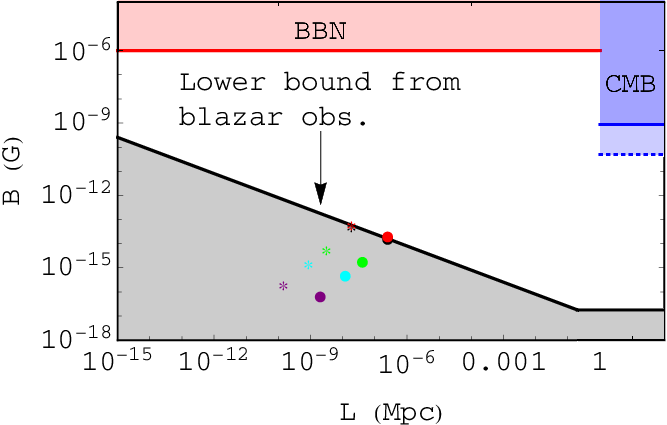}
\caption{Obtained magnetic field strength versus its coherence length for different values of $\alpha/f$.
The purple, cyan, green, black, and red correspond to $\alpha/f=35,50,60,75,90$, respectively. Bullet points, correspond to the case in which Alfv\'en time is considered as the decay timescale for the nonlinear evolution of the magnetic field after its generation. However, if one considers the reconnection time instead, the values change, and these are represented by stars in the corresponding colors. The black-shaded region denotes the lower bound on the inter-galactic magnetic field strength obtained from GeV observations of blazars \cite{MAGIC:2022piy}, while the red region represents the upper bound from Big Bang Nucleosynthesis (BBN) \cite{Kernan:1995bz}. The blue region indicates the upper bound obtained from Planck collaboration \cite{2016A&A...594A..19P} and blue dotted line represents the improved upper bound considering inflationary magnetic curvature mode in non-Gaussianity trispectrum \cite{Trivedi:2013wqa}.}
\label{estimate_magnetic_field}
\end{figure}

In the nonlinear processing regime, the magnetic field strength decays
as $(a_m/a_{nl})^{-1/3}$, and the coherence length (which corresponds
to the peak scale in our case) increases as $(a_m/a_{nl})^{2/3}$ for
the fully helical case.%
\footnote{As demonstrated in \App{scalar_spec} for $\alpha/f=75$ and $\alpha/f=90$, the relative magnetic helicity near the end of inflation depends on the wave number.
Notably, the field remains almost fully helical around the peak wave number of the power spectrum of the magnetic component.
Hence, a similar late-time evolution is expected as in the fully helical case. 
However, a detailed investigation would require MHD simulations in the radiation-dominated epoch for treating such a magnetic field energy density spectrum.}
By using $T=1 {\rm eV}$ (the temperature of the Universe at the matter-radiation equality) in \Eq{sf_ratio}, we obtain,
\begin{align}
\frac{a_{nl}}{a_m} = 1.13 \times 10^{-25} \left(\frac{g_{m,s}}{3.94} \frac{106.75}{g_{nl,s}}\right)^{1/3}\left(\frac{106.75}{g_{nl}}\right)^{1/4} \frac{T_m}{1 {\rm eV}} \left(\frac{ 10^{-6} m_{\rm pl}}{H}\right)^{1/2} r.
\end{align}
Here $r=\max(1,\mathcal{H}~L_c/V_A)$ and $T_m$ denotes the temperature of the Universe at the matter-radiation equality. We show the time evolution of $\mathcal{H}L_c/V_A$ in \App{NLP_time}
for the runs given in \Tab{table1}. After incorporating the nonlinear processing of the fields, the magnetic
field strength and its coherence length at the present epoch is given by
\begin{align}\label{mfstrength}
B_\mathrm{rms}\big|_0
&=9.2 \times 10^{-15}\, {\rm G}\,\sqrt{\frac{{\int d\log{k}\cdot P_{B}}}{\rho_e}}  \left(\frac{ 10^{-6} m_{\rm pl}}{H}\right)^{1/6}r^{1/3}
\end{align}and 
\begin{align}\label{cohlen}
L_c^\mathrm{NL}=0.8~ {\rm pc} ~(\mathcal{H}L_c)\left(\frac{10^{-6}m_{\rm pl}}{H}\right)^{1/6}\,r^{-2/3}.
\end{align}
After using the above expressions, we obtained the magnetic field strength
and its coherence length corresponding to each coupling strength and
the values are given in \Tab{table2}.
For the values given in columns 3 and 4 of \Tab{table2}, we use
$({\int d\log{k}\cdot P_{B}}/\rho)^{1/2}$ and $\mathcal{H}L_c$ at the end
of inflation in \Eq{mfstrength} and \Eq{cohlen}, respectively.
However, for the values given in columns 6 and 7, we use the value of
$({\int d\log{k}\cdot P_{B}}/\rho)^{1/2}$ and $\mathcal{H}L_c$ one $e$-fold
after inflation for the sake of comparison with the numbers in columns~2 and 3.
For run~E, we do not provide values at one $e$-fold after inflation,
as this would require a simulation with a resolution greater than $1024^3$.
In \Tab{table1}, $B_{\rm bound}$ denotes the lower bound of magnetic
field strength obtained from the blazars observation.
Here we assume $B_{\rm bound}=1.8 \times 10^{-17} {\rm G} \, (L_C^\mathrm{NL}/0.2 {\rm Mpc})^{1/2}$ \cite{MAGIC:2022piy}.
From the values given in the \Tab{table2}, we conclude that the magnetic
field strength obtained in the axion-U(1) inflation model for coupling
strengths $\alpha/f \ge 60$ can explain the required field strength to
account for the non-observation of GeV photons in blazar observations.
Recently, it has been suggested that the reconnection time,
rather than the Alfv\'en time, is the relevant timescale for the
decay of the nonlinear evolution of magnetic fields in the early
universe \cite{Zhou2019,Bhat2021,Hosking:2020wom,Hosking:2022umv}.
In \RRef{Brandenburg:2024tyi}, it was found that this makes the timescale
approximately 50 times longer than the Alfv\'en time.
If we account for this in our calculations, the length scale would
decrease by a factor of $50^{2/3}$, and the magnetic field strength
would increase by a factor of $50^{1/3}$ compared to the values mentioned
in \Tab{table2}.
(If the decay is controlled by reconnection and the decay time still
proportional to the square root of the conductivity, the exponents
might be 4/7 and 2/7, respectively; see \RRef{Schekochihin+22}.) 
In \Fig{estimate_magnetic_field}, we illustrate the obtained magnetic field strength and its coherence length for different values of $\alpha/f$, along with the lower and upper limits on the magnetic field strength from various observations. The purple, cyan, green, black, and red points correspond to $\alpha/f=35, 50, 60, 75$, and 90, respectively, in the case where the Alfv\'en time is taken as the decay time scale. The stars represent the corresponding values when the reconnection time is considered as the decay time scale.

In our study, we have not considered the production of fermions resulting from interactions
between gauge fields and fermions, e.g., via the Schwinger effect. This has been discussed in
refs.~\cite{Domcke:2018eki,Gorbar:2021rlt} for the case of a homogeneous axion field.
Upon incorporating the effects of charged currents generated by the
Schwinger effect, the strength of the produced magnetic fields
is expected to decrease by a factor of less than about 10, as can be seen
from figures~6 and 7 of \RRef{Gorbar:2021rlt}.
However, the exact reduction factor depends on the coupling strength.
The impact of this effect in the case of an inhomogeneous axion field is not
known, and we plan to address this in future work. In summary, we note that the gauge field
strengths presented in \Tab{table2} could be reduced if the effect of the produced charged currents were included.

\section{Curvature perturbations and its probability distribution function}\label{section5}

In the previous analytical study \cite{Linde2013,Bugaev:2013fya,Domcke:2017fix},
the formation of the PBHs from the large spatial fluctuations of the
axion field generated by the backreaction of the gauge field on the
axion dynamics has been studied.
To estimate the abundance of PBHs, the authors used the Press-Schechter
approach \cite{1974ApJ...187..425P,Anne2004,Byrnes:2012yx}.
For this, one needs to know the probability distribution function
(PDF) of the curvature perturbation variable,
$\zeta=-H\delta \phi/\dot{\bar{\phi}}$
\footnote{This expression for $\zeta$ is valid only within the linear approximation.}.
The PDF of $\zeta$ will be the same as the PDF of spatial fluctuations,
$\delta \phi$, of the axion field.
In these studies, this PDF has been assumed to be a $\chi^2$-distribution
due to the quadratic nature of the backreaction term of the gauge
field in the scalar field dynamics.
Using this, the authors in \cite{Linde2013,Bugaev:2013fya} examined the
constraints on $\alpha/f$ by using the constraints on PBHs abundances
from BBN and CMB and obtained an upper bound on $\alpha/f$.
The obtained values of the upper bound are
$\alpha/f \le 115$ from \cite{Linde2013} and
$\alpha/f \le 130$ from \cite{Bugaev:2013fya}. The obtained upper bound on $\alpha/f$ in these studies is such that this setup of axion-U(1) inflation is in the strong backreaction regime. A recent numerical study \cite{Caravano:2022epk} found that the non-Gaussianity of curvature perturbations is strongly suppressed, and the PDF of $\zeta$ is nearly Gaussian. Here we also test the assumption made in previous analytical studies \cite{Linde2013,Bugaev:2013fya} by analyzing the PDF of the axion field fluctuations in our simulations.

\begin{figure}[h!]
\centering
 \includegraphics[width=1\textwidth]{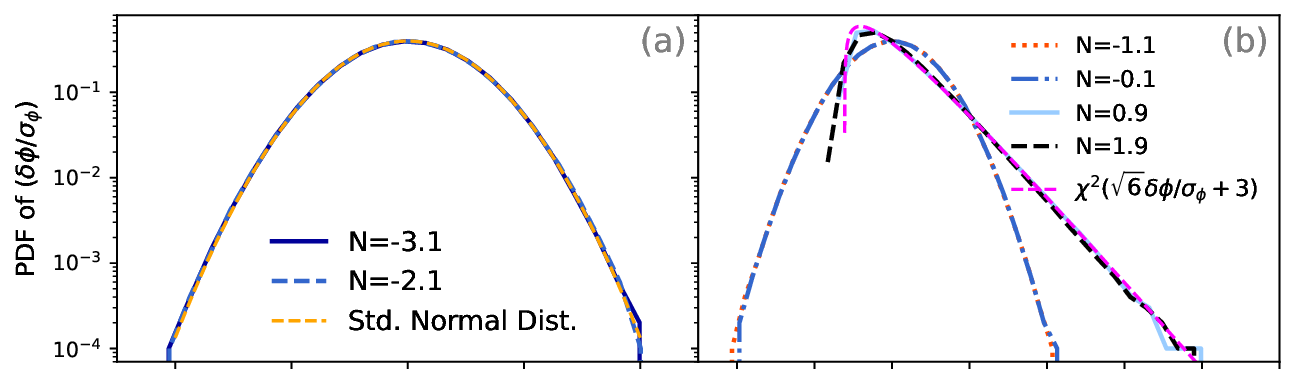}
  \includegraphics[width=1\textwidth]{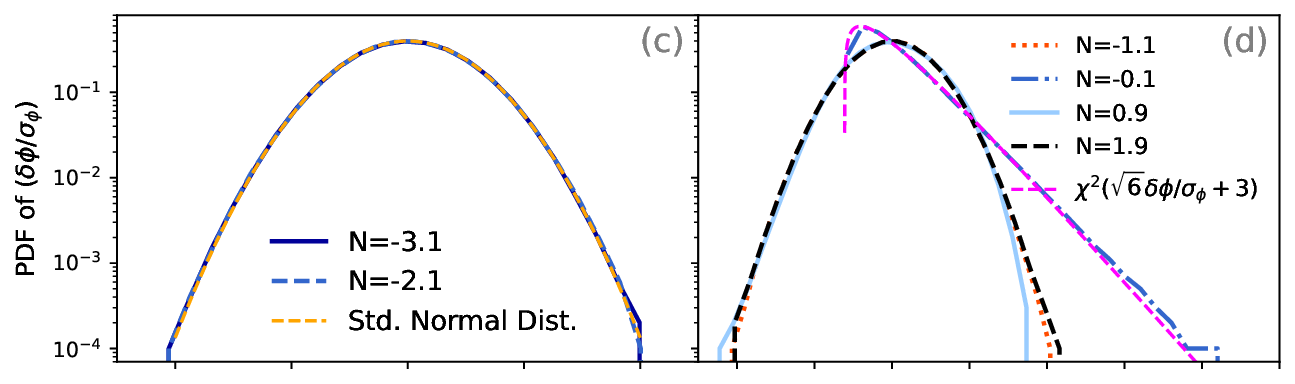}
 \includegraphics[width=1\textwidth]{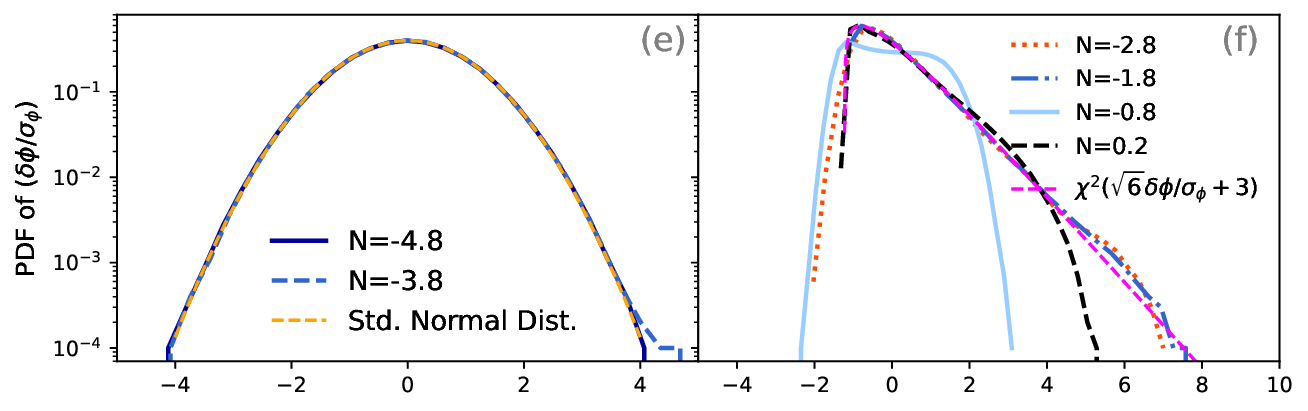}
\caption{The PDFs of $\delta\phi/\sigma_\phi$ at different times for
$\alpha/f=35$ (top row), $\alpha/f=60$ (middle row),
and $\alpha/f=75$ (bottom row).
The yellow line represents the PDF for a Gaussian random field, while
the magenta line shows the $\chi^2$-distribution with three degrees of
freedom for the variable $\sqrt{6}\delta\phi/\sigma_\phi+3$.
Here $N$ represents the number of $e$-folds and $N=0$ marks the end of inflation.
In the bottom row, the initial time corresponds to $N=-4.8$, i.e.,
$1.7$ $e$-folds less than that of the top and middle rows
to address the prolonged duration of inflation by
$1.7$ $e$-folds for the case of $\alpha/f=75$.
In the left column, we show the PDFs at the earlier times of the simulation, while the right column displays them at the later stages.
}
\label{pdf_of_deltaphi}
\end{figure}

To study the evolution of the PDF of the axion field's spatial
fluctuations, we use 3D snapshots from our simulations.
At the beginning of our simulations, these fluctuations follow a
Gaussian random field as set by the initial conditions.
We normalize by the root mean square of the fluctuations in the axion field, $\sigma_{\phi}$,
and plot the PDF of $\delta \phi/\sigma_\phi\equiv(\phi-\bar{\phi})/\sigma_{\phi}$.
Initially, it follows a standard normal distribution.
In \Fig{pdf_of_deltaphi}, we present the evolution of the PDFs for three
different coupling strengths, $\alpha/f=35$
(top panel), $\alpha/f=60$ (mid panel) and
$\alpha/f=75$ (bottom panel). In this figure, the dashed yellow line in the left column represents a standard normal distribution, while the dashed magenta line in the right column corresponds to a $\chi^2$-square distribution with three degrees of freedom. The variable $\delta \phi/\sigma_\phi$ has zero mean and unit variance. To compare it with a $\chi^2$-square distribution with different degrees of freedom, we use a transformed variable $\sqrt{2d}\,\delta \phi/\sigma_\phi+d$ with a mean of $d$ and a variance of $2d$. This transformation is used because a $\chi^2$-square distributed variable with $d$ degrees of freedom has a mean $d$ and a variance of $2d$. In \Fig{pdf_of_deltaphi}, we show the $\chi^2$-distribution for $d=3$, as it provides the best fit.
As is evident from the top panel of \Fig{pdf_of_deltaphi}, the
PDF of $\delta\phi/\sigma_\phi$ is initially Gaussian and evolves into
a $\chi^2$-distribution with three degrees of freedom.
However, in the case of strong coupling strength, the PDF starts as a
Gaussian, transitions to a $\chi^2$ distribution at an intermediate time,
and reverts to a distribution closer to Gaussian at later times. It is to be noted that the PDF at $N = 0.2$ for the
case $\alpha/f = 75$ has less power in the tails compared to the
$\chi^2$-distribution.
However, it is still not close to a Gaussian distribution. The exact
reason for this behavior is not clear, but it is expected that the PDF
at later times, $N > 0.2$, will have a more Gaussian-like distribution.

In a previous numerical study of axion-U(1) inflation
\cite{Caravano:2022epk}, the authors investigated the PDF of curvature
perturbations for $\alpha/f = 125$.
However, due to the limited box size, their simulations covered only 6 $e$-folds during inflation and did not extend until the end of inflation. Extending the simulation until the end would require a significantly larger simulation box. In their study, the authors also observed that the PDF tends toward a Gaussian distribution. We would like to emphasize, however, that this result is not exclusive to cases where the gauge field backreaction is significant during inflation. As demonstrated in the middle panel of \Fig{pdf_of_deltaphi}, the PDF of axion fluctuations for $\alpha/f = 60$ shows that this transition toward a Gaussian distribution also occurs when the backreaction becomes important after inflation.

Considering a distribution that is less non-Gaussian than the $\chi^2$-distribution will result in
a smaller PBH abundance\footnote{To accurately estimate PBH abundances, it is necessary to know the PDF of the smoothed curvature perturbations \cite{Linde2013,Bugaev:2013fya}. We expect that the PDF of the smoothed curvature perturbations will be similar to or even less non-Gaussian than that of the unsmoothed perturbations.
This is because smoothing involves averaging fluctuations using a window function that assigns different weights to points at different separations.
Hence, a non-Gaussian distribution will gradually approach a Gaussian one due to the central limit theorem.
Taking this into consideration, the constraints derived from the smoothed perturbations will be comparable to or less stringent than those from the unsmoothed perturbations.
}. This will relax the constraints on $\alpha/f$. We will present a detailed study of the PBHs constraints in a future study.

\section{Discussion and conclusion}\label{section6}

In this paper, we study the axion-U(1) inflation model by simulating the dynamics of both the axion and the gauge fields. Our focus is on the regime where the gauge field backreaction on axion evolution becomes significant during inflation. Previous studies have addressed this scenario semi-analytically using perturbative \cite{Domcke:2020zez,Domcke:2023tnn} and gradient expansion approaches \cite{Sobol:2020lec,Gorbar:2021rlt}, which neglect axion field inhomogeneity. However, recent numerical simulations have incorporated these inhomogeneities \cite{Caravano:2022epk,Figueroa:2023oxc,Caravano:2024xsb}. For our study, we developed two new modules in the {\sc Pencil Code}. Our numerical simulation results align well with previous semi-analytical studies for the homogeneous axion case and with recent numerical simulations for the inhomogeneous axion case.

We have analyzed the production of GWs and the PDF of spatial fluctuations in the axion field for $\alpha/f = 75$ and 90 in the Planck units.
We also compare these results with cases of smaller coupling strengths ($\alpha/f = 35$, $50$, and $60$), where the gauge field backreaction is not significant during inflation.
For $\alpha/f = 75$ and 90, the gauge field energy density becomes comparable to the potential energy of the axion field at the end of inflation and soon dominates the total energy budget of the universe. The produced GWs are within the bounds set by Planck data of CMB if only those produced until the end of inflation with no further production are considered. However, these GWs might exceed the bound if we account for those produced 2--3 $e$-folds during reheating after inflation and neglect the effects of the produced charged particles from the gauge fields. 

In this study, we have neglected the interaction of the gauge
field with other Standard Model particles.
In cases of large coupling, such as $\alpha/f = 75$ and 90, the gauge
field energy density is sufficient to provide instantaneous reheating,
as the interaction rate of the gauge field with other particles is
much larger than the Hubble expansion rate.
This is discussed in detail in \RRef{Adshead:2016iae}.
Additionally, the produced large gauge fields may lead to the production
of charged particles via the Schwinger effect \cite{PhysRev.82.664,Frob:2014zka,Kobayashi:2014zza,sharma2017,Domcke:2018eki,Sobol:2019xls,Fujita:2022fwc,vonEckardstein:2024tix}. Once the backreaction
of these charged particles becomes significant, they may halt any further
production of the gauge fields.
Consequently, the produced GW energy density will saturate at a stage
when these interactions become important. As these interactions may not be neglected until 2–3 $e$-folds after inflation,
the bound from the Planck data may not be exceeded as a result. A detailed study incorporating the effects of other interactions will be conducted in future work.

The large-scale gauge fields produced in this model offer a possible explanation for intergalactic magnetic fields. We have explored this for small to large coupling strengths, $\alpha/f$, and find that the resulting magnetic fields are only marginally above the lower bound on intergalactic magnetic fields inferred from blazar observations \cite{Neronov:1900zz} for $\alpha/f \ge 60$.
For the coupling strength $\alpha/f=90$, the resulting magnetic field
strength is $1.9 \times 10^{-14}\G$ with a coherence length of $0.25\pc$.
The lower bound on the magnetic field strength from blazar observations
for this coherence length is $1.6 \times 10^{-14}\G$, based on the
relation $B_{\rm bound} =1.8 \times 10^{-17} \G \left(L / 0.2 \Mpc\right)^{1/2}$,
as discussed in \Sec{NLP_section}.
We note that the generated magnetic field strengths could be reduced upon incorporating the effects of charged currents produced through interactions between gauge fields and fermions, e.g.,
via the Schwinger effect. We leave a detailed investigation of this effect for future work.

Furthermore, we examine the PDF of the produced axion fluctuations. Understanding this PDF is crucial for studying the formation of PBHs in this model. Previous analytical studies considered it to be a $\chi^2$-distribution with one or two degrees of freedom. In our study, we initialize axion fluctuations as a Gaussian random field. We find that the spatial fluctuations indeed transition to a $\chi^2$-distribution at later times, but with three degrees of freedom. However, as the backreaction of the gauge field becomes significant, even if only after inflation, the system starts to resemble a Gaussian distribution as also previously demonstrated in \RRef{Caravano:2022epk} for the case $\alpha/f=125$, when the backreaction is important already during inflation.
Incorporating this information in studies constraining the axion-gauge coupling from PBH abundance bounds will relax the existing constraints on this coupling strength. 

In this study, we examine the unsmoothed axion fluctuation distribution. However, to make accurate estimates, one needs to understand the distribution of the smoothed fluctuations. We believe that the distribution of smoothed fluctuations will be even closer to a Gaussian distribution compared to the unsmoothed case. We will provide a detailed study on this in a future work.

\section*{Acknowledgments}
We would like to thank Angelo Caravano, Valerie Domcke,
Yutong He, Oksana Iarygina, and Alberto Roper Pol for the discussions on this project.
We are also indebted to Kai Schmitz and Oleksandr Sobol for providing
the datasets used to compare with the results from the gradient expansion
formalism for the homogeneous axion case
as well as for the discussions on this comparison.
R.S.\ and A.V.\ were supported by the Czech Science Foundation (GA\v{C}R),
project 24-13079S.
A.B.\ acknowledges funding from the Swedish Research Council
(Vetenskapsr{\aa}det) under grant No.\ 2019-04234 and the National
Science Foundation under Grant No.\ AST-2307698,
and the NASA ATP Award 80NSSC22K0825.
We also thank the Swedish National Allocations Committee for providing
computing resources at the Center for Parallel Computers at the Royal
Institute of Technology in Stockholm and the National Supercomputer Centre
(NSC) in Link\"oping. Some simulations in this work were performed on
the ``Phoebe'' computer cluster at CEICO/FZU. R.S. is grateful to Josef
Dvo\v{r}\'a\v{c}ek
for assistance with setting up the \textsc{Pencil Code} on Phoebe.
K.S.\ was partially supported by the Alexander Von Humboldt Foundation through the
Carl Friedrich von Siemens Research Award. He thanks Volker Springel his host, as well as
Torsten Ensslin and Eichiro Komatsu for warm hospitality
during his stay at the Max-Planck Institute for Astrophysics, Garching.

\paragraph{Data availability.}
The source code used for the numerical solutions of
this study, the \textsc{Pencil Code}, along with the module
\texttt{special/backreact\_infl} used in the present study, are freely
available at \url{https://github.com/pencil-code/pencil-code/}.
The numerical data and input files are available on Zenodo; see ref.~\cite{DATA}.

\appendix
\section{Implementation in the \textsc{Pencil Code}}\label{AppendixA}

As discussed in \Sec{section1}, we use the \textsc{Pencil Code} for
our simulations.
To simulate the axion-U(1) inflation, we use the special modules
\texttt{backreact\_inf.f90} and \texttt{disp\_current.f90}.
The $\phi$ and FLRW background equations are solved in the
`backreact\_infl.f90' module, while the $\EE$ and $\AA$ equations are
solved in the `disp\_current.f90' module.
By default, we use the Weyl gauge, but the Lorenz gauge can also
be used by setting the \texttt{llongitudinalE} logical to false and the
\texttt{llorenz\_gauge\_disp} logical to true.
The governing equations in both Lorenz and Weyl gauges are described
in the \Sec{lgauge} and \Sec{wgauge}, respectively.
In addition to the $\EE$ and $\AA$ variables, we also solve for $A_0$
in the Lorenz gauge.
Additionally, we also solve for the production of GWs
using the equations in the gravitational module \texttt{gravitational\_waves\_hTXk.f90};
see \RRef{RoperPol+20} for details.
The \textsc{Pencil Code} employs a sixth-order finite difference method
for spatial derivatives and a third-order Runge-Kutta method for time
stepping.

\subsection{Lorenz Gauge}\label{lgauge}

In the Lorenz gauge, where $A_0'=\Del \cdot \AA$, the system of equations
\eqref{phi_eqn}, \eqref{A0_eqn}, \eqref{Ai_eqn} and \eqref{hequation}
reduces to
\begin{align}
\phi''+2\hhh\phi'-\ddel \phi+a^2\pfrac{V}{\phi}&=\frac{\alpha}{f}a^{-2}\,\EE\cdot \BB \label{phieq}\\
A_0'&=\deldot \AA\quad \text{ (Lorenz gauge)}\quad \label{lgeq}\\
A_0''-\ddel A_0=\frac{\alpha}{f}\del \phi \cdot \BB \quad &\text{or} \quad (\deldot \AA)'-\ddel A_0=\frac{\alpha}{f}\del \phi \cdot \BB \quad \label{deldotaeq}\\
\AA'&=-\EE+\del A_0 \label{aeq}\\
\EE'&=\delcrs \BB -\frac{\alpha}{f}\left(\phi' \BB+\del \phi \times\EE\right). \label{efeq}\\
     \hhh^2&=\frac{8\pi}{3 m_\mathrm{pl}^2}a^2 \rho \label{hubbleeq}.
\end{align}
Here $\EE=\del A_0-\AA'$ and $\BB=\delcrs \AA$. 
In this case, we solve the equations \eqref{phieq}, \eqref{lgeq},
\eqref{deldotaeq}, \eqref{aeq} and \eqref{efeq} for the axion-gauge field dynamics and \Eq{hubbleeq} for the FLRW background evolution.

\subsection{Weyl or temporal gauge}\label{wgauge}

In the Weyl or temporal gauge, we have $A_0=0$ and instead of equations
\eqref{lgeq}, \eqref{deldotaeq}, \eqref{aeq}, and \eqref{efeq}, we solve
equations \eqref{aeq1} and \eqref{efeq1} and \eqref{gammaeq} for the
gauge field evolution.
\begin{align}
\AA'&=-\EE\label{aeq1}\\
\EE'&=-\ddel \AA+\del \Gamma -\frac{\alpha}{f}\left(\phi' \BB+\del \phi \times\EE\right)\label{efeq1} \\ 
\Gamma'&= -(1-g_e^2)\deldot \EE+g_e^2 \frac{\alpha}{f}\del \phi \cdot \BB. \label{gammaeq}
\end{align}
Here $\Gamma=\deldot \AA$.
In \App{AdditionalTests}, we present a comparison of the different gauges.

\section{Different initial cutoff wave numbers}\label{kp_test}

As mentioned in \Sec{InitialConditions}, we initialize the fluctuations of
the axion and gauge fields using the Bunch-Davies initial condition.
To neglect the contribution of large ultraviolet quantum fluctuation
modes, we consider an initial power spectrum with a cutoff at a
wave number $k_p$, above which the power spectrum follows a decaying
power law.
In this appendix, we demonstrate that the results of our simulations
are independent of the initial value of $k_p$.

\begin{figure}[h!]
\centering
 \includegraphics[width=1\textwidth]{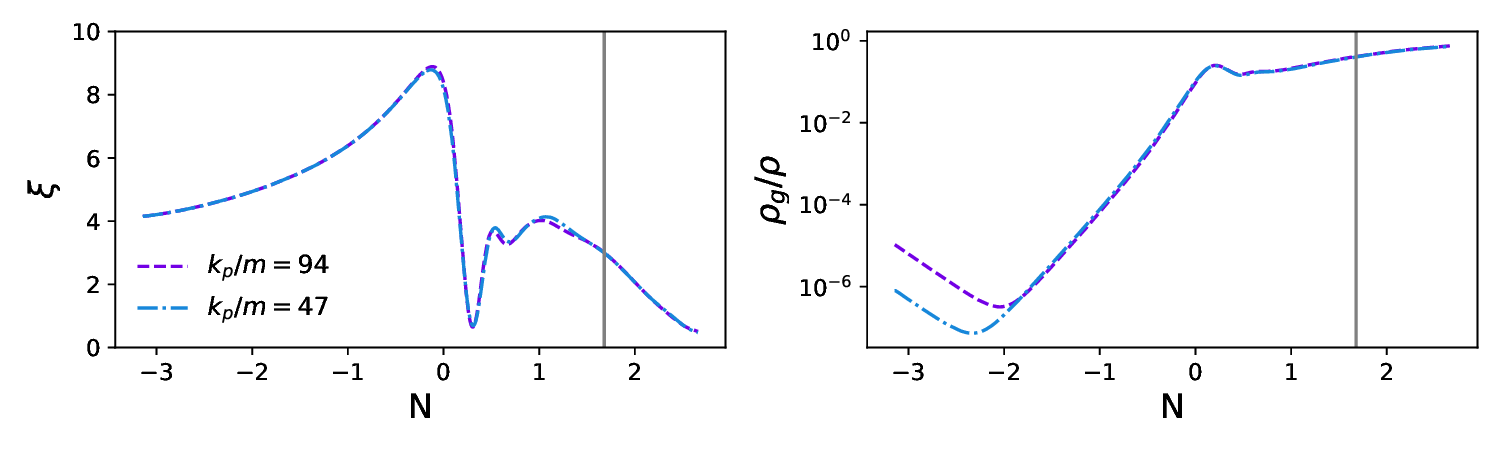}
\caption{In this figure, we show the evolution of $\xi$ and $\rho_g/\rho$ for the runs D and its analogues run in which $k_p/m=94$.}
\label{different_resolution}
\end{figure}

We conduct a simulation analogous to run~D with $k_p/m=94$ and $\alpha/f=75$.
The results are shown in \Fig{different_resolution}, which depicts
the evolution of $\xi$ defined below \Eq{Aequation_woback} and the
ratio of the gauge field energy density to the total energy density
$\rho_g/\rho$ for $k_p/m=47$ (dot-dashed blue curves) and $k_p/m=94$
(dashed violet curves).
As evident from \Fig{different_resolution}, both sets of curves overlap,
indicating that the location of $k_p$ does not affect the final results.
The initial departure in the two curves in the right panel is because
of the larger value of the initial amplitude of the gauge fields for
the case $k_p/m=94$.
We also checked the dependence on the cutoff wave number for $\alpha/f=90$ and came to similar conclusions for this case as well.

\section{Comparison with the GEF results}\label{app_comp_with_gef}

\begin{figure}[h!]
\centering
 \includegraphics[width=1\textwidth]{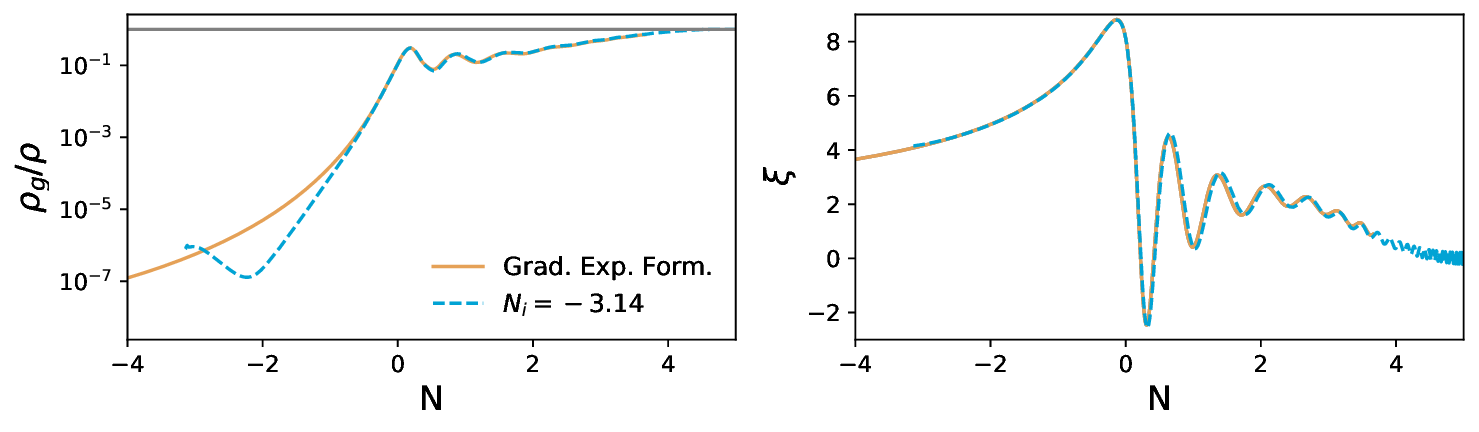}
 \includegraphics[width=1\textwidth]{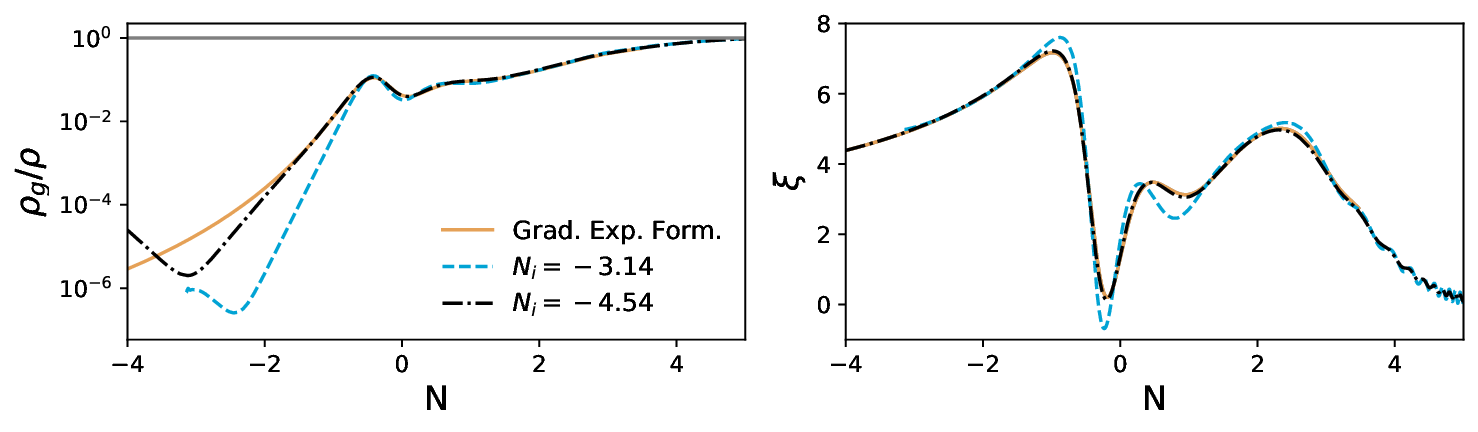}
\caption{The evolution of the ratio of gauge field energy density to
the total energy density (left panel) and the parameter $\xi$ is shown
in this figure.
The yellow curves represent the results from the GEF approach, while
the blue and black curves show results from our simulation for the
homogeneous axion case.
The black curve corresponds to the case where we start at $N_i = -4.54$
$e$-folds before the end of slow-roll inflation, and the blue curve is
for $N_i = -3.14$.
The upper and lower rows are for the coupling strength, $\alpha=75$
and $90$, respectively.}
\label{comp_with_gef}
\end{figure}

In \Sec{ComparisonWithPreviousWork}, we compared with the GEF results.
Here, we discuss the need to choose initial conditions in the simulations
that allow the system to relax into the slow-roll regime a few $e$-folds
before the gauge field backreaction becomes significant.
In our simulation, we initialize $\phi$ and $\phi'$ based on slow-roll
conditions, starting the simulation $3.14$ $e$-folds before the end of
slow-roll inflation when gauge field backreaction is not yet significant.
The spectra of $\AA$ and $\EE$ are initialized using the Bunch-Davies conditions.
However, gauge modes closer to the horizon may already have been
amplified, i.e., the Bunch-Davies conditions might not be an ideal
initial condition for all gauge modes.
We expect the system to naturally evolve to these solutions within a
few $e$-folds, as can be seen for the coupling strength $\alpha/f=75$
in \Fig{comp_with_gef}, where this alignment occurs about one $e$-folds
before the end of inflation, yielding a good match between GEF and our
simulation results.

This is to be noted that, for $\alpha/f=90$, we initialize the system at
$N_i = -4.5$ to provide more time for it to approach the GEF results.
If we instead initialize this case at $N_i = -3.14$, the gauge field
energy density agrees with the GEF results only just before the gauge
fields' backreaction becomes significant for the axion dynamics, as
shown by the blue curve in \Fig{comp_with_gef}.
This initialization does not provide a good match for the evolution
of $\xi$ at later times, as shown in the bottom right panel in
\Fig{comp_with_gef}.

Importantly, the difference in gauge field energy density between GEF
and our simulation remains small after the time of backreaction even in
the case in which we initialize at $N_i=-3.14$ for $\alpha/f=90$.
Therefore, while choosing better initial conditions could improve
consistency with GEF, it would not significantly impact the predictions
for magnetic fields and GWs.

\section{Spectrum of scalar field fluctuations and of helicity of the magnetic component of the gauge field
}\label{scalar_spec}

\begin{figure}[h!]
\centering
 \includegraphics[width=1\textwidth]{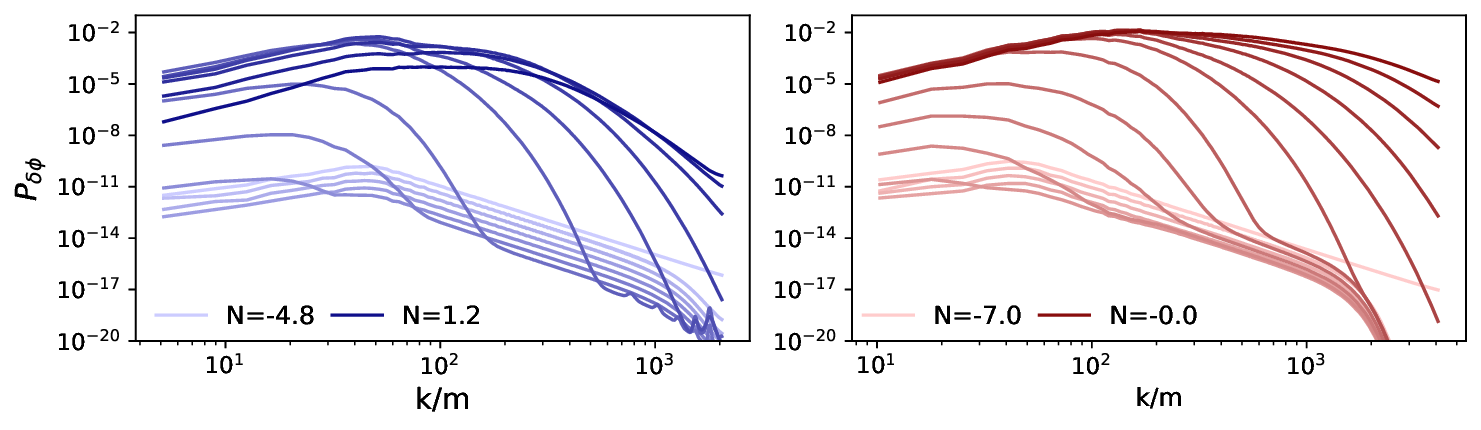}
\includegraphics[width=1\textwidth]{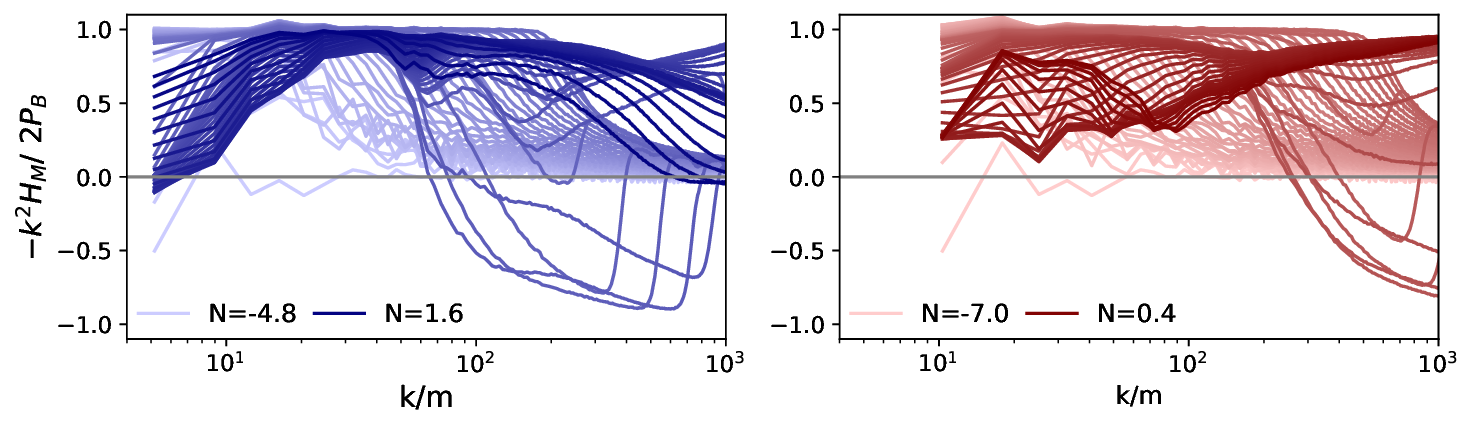}
\caption{The spectrum of scalar field fluctuations
and of the helicity of the magnetic component of the gauge field
is shown, with the left panel representing run D and the right panel representing run E.
The curves progress in time from lighter to darker shades, each separated by $\Delta N = 0.5$ for the upper panel and $\Delta N=0.1$ for the lower panel.}
\label{scalar_spectra}
\end{figure}

At the end of \Sec{ComparisonWithPreviousWork}, we compared with the
results for runs~D and E.
Here, we provide the spectra of the scalar field fluctuation $P_{\delta \phi}$ and the spectra of the helicity of the magnetic component of the gauge field, $H_M(k)$ defined as $\int {\rm d }k~H_M(k)=\langle (\AA \cdot \BB) \rangle$.
The left- and right-hand panels of \Fig{scalar_spectra} show those
spectra for runs D and E, respectively.
The light to dark shades represent different times, with each curve
separated by $\Delta N = 0.5$.
They agree well with those of \RRef{Figueroa:2023oxc}.

\section{Time evolution of $\mathcal{H}~L_c/V_A$}\label{NLP_time}

\begin{figure}[h!]
\centering
 \includegraphics[width=0.8\textwidth]{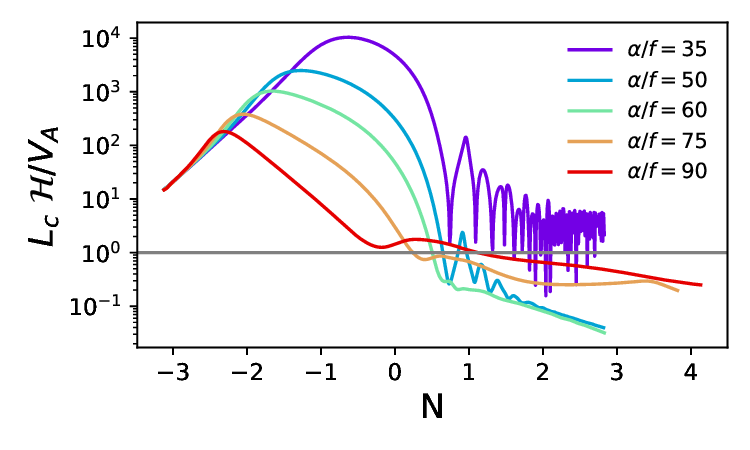}
\caption{Time evolution of $\mathcal{H}~L_c/V_A$ for the runs given in \Tab{table1}.}
\label{value of r}
\end{figure}

In \Sec{NLP_section}, we discussed that the nonlinear processing of the
magnetic field depends on the ratio of the Alfv\'en time scale to the
Hubble time scale, $\mathcal{H}~L_c/V_A$.
This ratio shown in \Fig{value of r} decides when the nonlinear processing of the magnetic field will begin in the plasma after its generation.
The different curves show the time evolution of $\mathcal{H}~L_c/V_A$
for different values of $\alpha/f$.

From this figure, we see that after the end of inflation, the value of $\mathcal{H}~L_c/V_A$ is less than 1 for all cases except for $\alpha/f = 35$.
Therefore, as discussed in \Sec{NLP_section}, we conclude that the magnetic fields are in the nonlinear processing regime at the end of inflation, except in the case of $\alpha/f = 35$.

\section{Additional tests}
\label{AdditionalTests}

In \App{AppendixA}, we stated that we use the Weyl gauge by default.
Here, we compare with the Lorenz gauge and show that with both gauges,
the constraint \Eq{A0_eqn} is obeyed.

\subsection{Weyl vs Lorenz gauge}\label{weyl_and_lorenz_g}

We use the Weyl gauge to evolve the gauge field in the simulations discussed in this paper.
However, the Lorenz gauge can also be used to solve these equations, although this requires solving an additional equation—the evolution of the temporal component, $A_0$.
The Lorenz gauge has been used in previous studies \cite{Adshead:2016iae,Caravano:2022epk}.

\begin{figure}[h!]
\centering
 \includegraphics[width=1\textwidth]{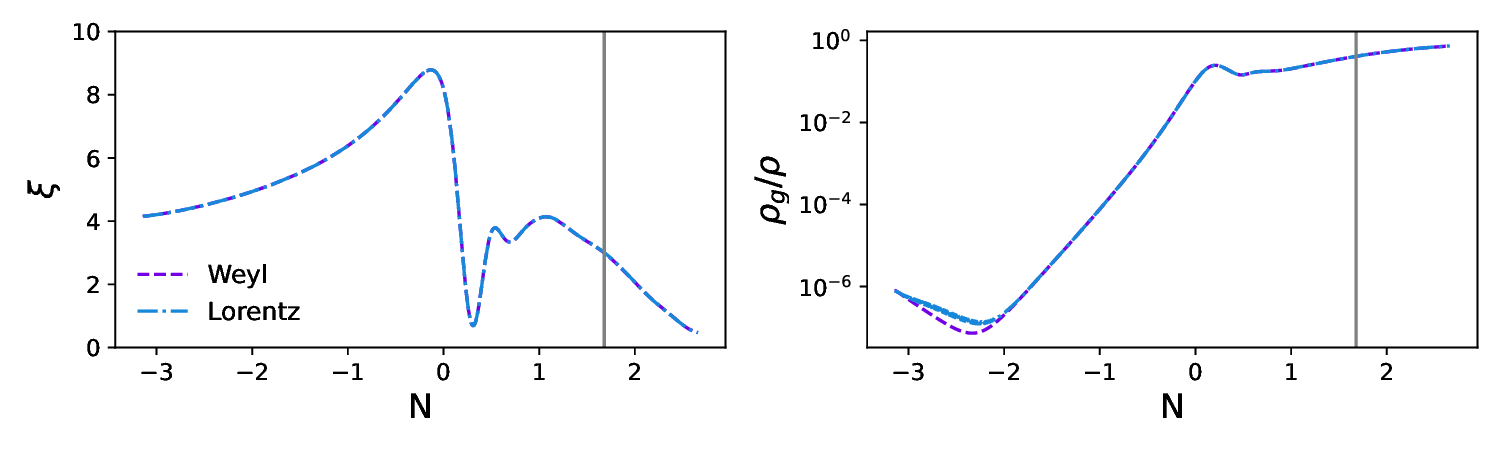}
\caption{In this figure, we show the evolution of $\xi$ and the ratio $\rho_g/\rho$ for run~D (dashed violet curves),
along with its analogous run where the gauge field equation is evolved using the Lorenz gauge (dashed-dotted blue curves).}
\label{lorenz_vs_weyl}
\end{figure}

In this section, we compare the results of run D for the Weyl and
Lorenz gauges.
Thus, we run a simulation analogous to run D in the Lorenz gauge, using
\Eq{lgauge}, and show the time evolution of the parameter $\xi$ and the
ratio $\rho_g/\rho$ in \Fig{lorenz_vs_weyl}.
From this figure, it is clear that the evolution in the Lorenz gauge
(dashed-dotted blue curves) closely matches that in the Weyl gauge (dashed
violet curves).
Thus, either gauge can be used for the numerical evolution of the gauge
field equations.

\subsection{Constraint equation check}

As discussed in \App{AppendixA}, with the {\sc Pencil Code}, the
axion-U(1) inflation model can be solved using either the Lorenz gauge
or the Weyl gauge.
While the simulations discussed in the main text use the Weyl gauge,
we provide a comparison of the results obtained with both gauges in
the \App{weyl_and_lorenz_g}.
Here, we demonstrate the status of the constraint \Eq{A0_eqn}
with time in one of our simulations.
To do this, we use the following expression
\begin{align}
C.E.=\Big\langle \frac{\deldot\EE-\frac{\alpha}{f}\Del \phi \cdot \BB}{\sqrt{(\deldot\EE)^2+(\frac{\alpha}{f}\Del \phi \cdot \BB)^2}}\Big\rangle.
\end{align}
\begin{figure}[h!]
\centering
 \includegraphics[width=1\textwidth]{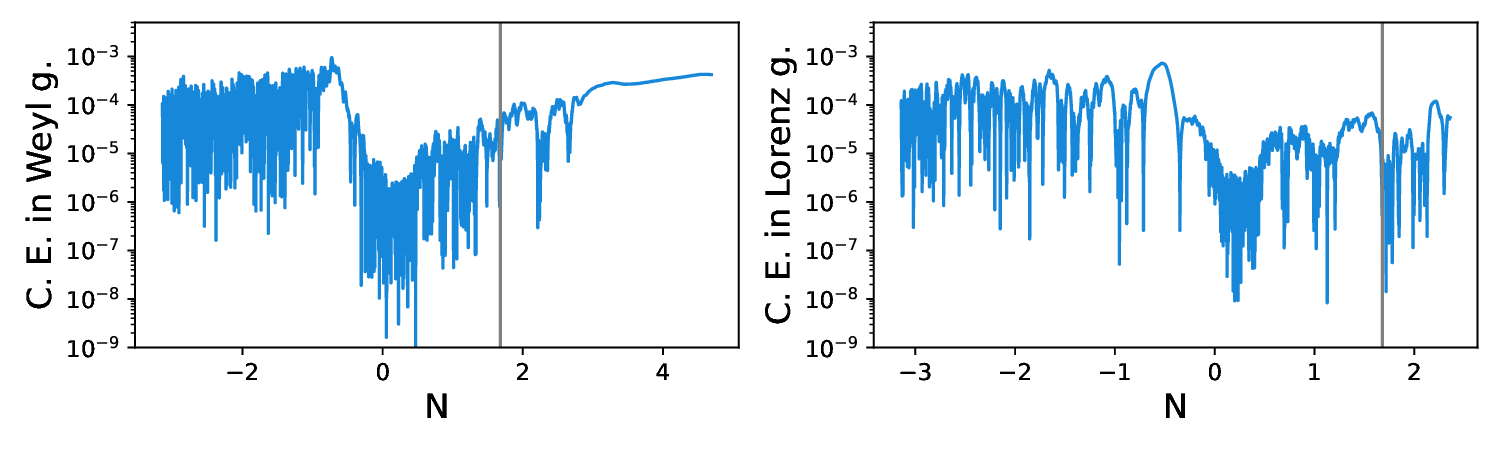}
\caption{The status of the constraint \Eq{A0_eqn}
with time in Weyl (left panel) and Lorenz gauge (right panel) is shown for the run D.}
\label{constraints_eqn_check}
\end{figure}
In \Fig{constraints_eqn_check}, the left panel shows the status of the
constraint equation in the Weyl gauge, while the right panel shows the
gauge condition in the Lorenz gauge.
Both panels correspond to run D. 
Additionally, we also check the accuracy with which the constraint equation is satisfied for different values of $g_e^2$ defined in \Eq{gammaeq}.
We find that for $g_e^2=2$, $1$, or $0$, the accuracy remains at similar levels shown in \Fig{constraints_eqn_check}.

\bibliographystyle{bibi}
\bibliography{references}

\providecommand{\href}[2]{#2}\begingroup\raggedright\begin{thebibliography}{100}

\bibitem{STAROBINSKY198099}
A.~Starobinsky, \emph{A new type of isotropic cosmological models without
  singularity},
  \href{https://doi.org/https://doi.org/10.1016/0370-2693(80)90670-X}{\emph{Phys.
  Lett. B} {\bfseries 91} (1980) 99}.

\bibitem{Guth1981}
A.~H. Guth, \emph{Inflationary universe: A possible solution to the horizon and
  flatness problems},
  \href{https://doi.org/10.1103/PhysRevD.23.347}{\emph{Phys. Rev. D} {\bfseries
  23} (1981) 347}.

\bibitem{LINDE1982389}
A.~Linde, \emph{A new inflationary universe scenario: A possible solution of
  the horizon, flatness, homogeneity, isotropy and primordial monopole
  problems},
  \href{https://doi.org/https://doi.org/10.1016/0370-2693(82)91219-9}{\emph{Phys.
  Lett. B} {\bfseries 108} (1982) 389}.

\bibitem{PhysRevLett.48.1220}
A.~Albrecht and P.~J. Steinhardt, \emph{Cosmology for grand unified theories
  with radiatively induced symmetry breaking},
  \href{https://doi.org/10.1103/PhysRevLett.48.1220}{\emph{Phys. Rev. Lett.}
  {\bfseries 48} (1982) 1220}.

\bibitem{Kazanas1980}
D.~{Kazanas}, \emph{{Dynamics of the universe and spontaneous symmetry
  breaking}}, \href{https://doi.org/10.1086/183361}{\emph{Astrophys. J. Lett.}
  {\bfseries 241} (1980) L59}.

\bibitem{SATO1981311}
K.~Sato, \emph{The first order phase transition of a vacuum and baryon-number
  domain structure of the universe},
  \href{https://doi.org/https://doi.org/10.1016/0146-6410(81)90039-9}{\emph{Prog.
  Part. Nucl. Phys.} {\bfseries 6} (1981) 311}.

\bibitem{Mukhanov:1981xt}
V.~F. Mukhanov and G.~V. Chibisov, \emph{{Quantum Fluctuations and a
  Nonsingular Universe}}, {\emph{JETP Lett.} {\bfseries 33} (1981) 532}.

\bibitem{Chibisov:1982nx}
G.~V. Chibisov and V.~F. Mukhanov, \emph{{Galaxy formation and phonons}},
  {\emph{Mon. Not. Roy. Astron. Soc.} {\bfseries 200} (1982) 535}.

\bibitem{Mukhanov:2013tua}
V.~Mukhanov, \emph{{Quantum Cosmological Perturbations: Predictions and
  Observations}},
  \href{https://doi.org/10.1140/epjc/s10052-013-2486-7}{\emph{Eur. Phys. J. C}
  {\bfseries 73} (2013) 2486}
  [\href{https://arxiv.org/abs/1303.3925}{{\ttfamily 1303.3925}}].

\bibitem{Starobinsky:1982ee}
A.~A. Starobinsky, \emph{{Dynamics of Phase Transition in the New Inflationary
  Universe Scenario and Generation of Perturbations}},
  \href{https://doi.org/10.1016/0370-2693(82)90541-X}{\emph{Phys. Lett. B}
  {\bfseries 117} (1982) 175}.

\bibitem{Guth:1982ec}
A.~H. Guth and S.~Y. Pi, \emph{{Fluctuations in the New Inflationary
  Universe}}, \href{https://doi.org/10.1103/PhysRevLett.49.1110}{\emph{Phys.
  Rev. Lett.} {\bfseries 49} (1982) 1110}.

\bibitem{HAWKING1982295}
S.~Hawking, \emph{The development of irregularities in a single bubble
  inflationary universe},
  \href{https://doi.org/https://doi.org/10.1016/0370-2693(82)90373-2}{\emph{Phys.
  Lett. B} {\bfseries 115} (1982) 295}.

\bibitem{Bardeen:1983qw}
J.~M. Bardeen, P.~J. Steinhardt and M.~S. Turner, \emph{{Spontaneous Creation
  of Almost Scale - Free Density Perturbations in an Inflationary Universe}},
  \href{https://doi.org/10.1103/PhysRevD.28.679}{\emph{Phys. Rev. D} {\bfseries
  28} (1983) 679}.

\bibitem{Planck:2018nkj}
{\scshape Planck} Collaboration, N.~Aghanim et~al., \emph{{Planck 2018 results.
  I. Overview and the cosmological legacy of Planck}},
  \href{https://doi.org/10.1051/0004-6361/201833880}{\emph{Astron. Astrophys.}
  {\bfseries 641} (2020) A1}
  [\href{https://arxiv.org/abs/1807.06205}{{\ttfamily 1807.06205}}].

\bibitem{Baumann:2009ds}
D.~Baumann, \emph{{Inflation}},  in \emph{{Theoretical Advanced Study Institute
  in Elementary Particle Physics}: {Physics of the Large and the Small}},
  pp.~523--686, 2011, \href{https://arxiv.org/abs/0907.5424}{{\ttfamily
  0907.5424}}, \href{https://doi.org/10.1142/9789814327183_0010}{DOI}.

\bibitem{Sriramkumar:2009kg}
L.~Sriramkumar, \emph{{An introduction to inflation and cosmological
  perturbation theory}}, {\emph{Curr. Sci.} {\bfseries 97} (2009) 868}
  [\href{https://arxiv.org/abs/0904.4584}{{\ttfamily 0904.4584}}].

\bibitem{Martin:2013tda}
J.~Martin, C.~Ringeval and V.~Vennin, \emph{{Encyclop\ae{}dia Inflationaris}},
  \href{https://doi.org/10.1016/j.dark.2014.01.003}{\emph{Phys. Dark Univ.}
  {\bfseries 5-6} (2014) 75} [\href{https://arxiv.org/abs/1303.3787}{{\ttfamily
  1303.3787}}].

\bibitem{Planck:2018jri}
{\scshape Planck} Collaboration, Y.~Akrami et~al., \emph{{Planck 2018 results.
  X. Constraints on inflation}},
  \href{https://doi.org/10.1051/0004-6361/201833887}{\emph{Astron. Astrophys.}
  {\bfseries 641} (2020) A10}
  [\href{https://arxiv.org/abs/1807.06211}{{\ttfamily 1807.06211}}].

\bibitem{Chowdhury:2019otk}
D.~Chowdhury, J.~Martin, C.~Ringeval and V.~Vennin, \emph{{Assessing the
  scientific status of inflation after Planck}},
  \href{https://doi.org/10.1103/PhysRevD.100.083537}{\emph{Phys. Rev. D}
  {\bfseries 100} (2019) 083537}
  [\href{https://arxiv.org/abs/1902.03951}{{\ttfamily 1902.03951}}].

\bibitem{Freese1990}
K.~Freese, J.~A. Frieman and A.~V. Olinto, \emph{Natural inflation with pseudo
  nambu-goldstone bosons},
  \href{https://doi.org/10.1103/PhysRevLett.65.3233}{\emph{Phys. Rev. Lett.}
  {\bfseries 65} (1990) 3233}.

\bibitem{Adams:1992bn}
F.~C. Adams, J.~R. Bond, K.~Freese, J.~A. Frieman and A.~V. Olinto,
  \emph{{Natural inflation: Particle physics models, power law spectra for
  large scale structure, and constraints from COBE}},
  \href{https://doi.org/10.1103/PhysRevD.47.426}{\emph{Phys. Rev. D} {\bfseries
  47} (1993) 426} [\href{https://arxiv.org/abs/hep-ph/9207245}{{\ttfamily
  hep-ph/9207245}}].

\bibitem{McAllister:2008hb}
L.~McAllister, E.~Silverstein and A.~Westphal, \emph{{Gravity Waves and Linear
  Inflation from Axion Monodromy}},
  \href{https://doi.org/10.1103/PhysRevD.82.046003}{\emph{Phys. Rev. D}
  {\bfseries 82} (2010) 046003}
  [\href{https://arxiv.org/abs/0808.0706}{{\ttfamily 0808.0706}}].

\bibitem{Kaloper:2008fb}
N.~Kaloper and L.~Sorbo, \emph{{A Natural Framework for Chaotic Inflation}},
  \href{https://doi.org/10.1103/PhysRevLett.102.121301}{\emph{Phys. Rev. Lett.}
  {\bfseries 102} (2009) 121301}
  [\href{https://arxiv.org/abs/0811.1989}{{\ttfamily 0811.1989}}].

\bibitem{Kim:2004rp}
J.~E. Kim, H.~P. Nilles and M.~Peloso, \emph{{Completing natural inflation}},
  \href{https://doi.org/10.1088/1475-7516/2005/01/005}{\emph{JCAP} {\bfseries
  01} (2005) 005} [\href{https://arxiv.org/abs/hep-ph/0409138}{{\ttfamily
  hep-ph/0409138}}].

\bibitem{Anber:2009ua}
M.~M. Anber and L.~Sorbo, \emph{{Naturally inflating on steep potentials
  through electromagnetic dissipation}},
  \href{https://doi.org/10.1103/PhysRevD.81.043534}{\emph{Phys. Rev. D}
  {\bfseries 81} (2010) 043534}
  [\href{https://arxiv.org/abs/0908.4089}{{\ttfamily 0908.4089}}].

\bibitem{Barnaby:2010vf}
N.~Barnaby and M.~Peloso, \emph{{Large Nongaussianity in Axion Inflation}},
  \href{https://doi.org/10.1103/PhysRevLett.106.181301}{\emph{Phys. Rev. Lett.}
  {\bfseries 106} (2011) 181301}
  [\href{https://arxiv.org/abs/1011.1500}{{\ttfamily 1011.1500}}].

\bibitem{Barnaby:2011qe}
N.~Barnaby, E.~Pajer and M.~Peloso, \emph{{Gauge Field Production in Axion
  Inflation: Consequences for Monodromy, non-Gaussianity in the CMB, and
  Gravitational Waves at Interferometers}},
  \href{https://doi.org/10.1103/PhysRevD.85.023525}{\emph{Phys. Rev. D}
  {\bfseries 85} (2012) 023525}
  [\href{https://arxiv.org/abs/1110.3327}{{\ttfamily 1110.3327}}].

\bibitem{Adshead:2012kp}
P.~Adshead and M.~Wyman, \emph{{Chromo-Natural Inflation: Natural inflation on
  a steep potential with classical non-Abelian gauge fields}},
  \href{https://doi.org/10.1103/PhysRevLett.108.261302}{\emph{Phys. Rev. Lett.}
  {\bfseries 108} (2012) 261302}
  [\href{https://arxiv.org/abs/1202.2366}{{\ttfamily 1202.2366}}].

\bibitem{Notari:2016npn}
A.~Notari and K.~Tywoniuk, \emph{{Dissipative Axial Inflation}},
  \href{https://doi.org/10.1088/1475-7516/2016/12/038}{\emph{JCAP} {\bfseries
  12} (2016) 038} [\href{https://arxiv.org/abs/1608.06223}{{\ttfamily
  1608.06223}}].

\bibitem{Ferreira:2017lnd}
R.~Z. Ferreira and A.~Notari, \emph{{Thermalized Axion Inflation}},
  \href{https://doi.org/10.1088/1475-7516/2017/09/007}{\emph{JCAP} {\bfseries
  09} (2017) 007} [\href{https://arxiv.org/abs/1706.00373}{{\ttfamily
  1706.00373}}].

\bibitem{Pajer:2013fsa}
E.~Pajer and M.~Peloso, \emph{{A review of Axion Inflation in the era of
  Planck}}, \href{https://doi.org/10.1088/0264-9381/30/21/214002}{\emph{Class.
  Quant. Grav.} {\bfseries 30} (2013) 214002}
  [\href{https://arxiv.org/abs/1305.3557}{{\ttfamily 1305.3557}}].

\bibitem{Alam:2024fid}
K.~Alam, K.~Dutta and N.~Jaman, \emph{{CMB constraints on natural inflation
  with gauge field production}},
  \href{https://doi.org/10.1088/1475-7516/2024/12/015}{\emph{JCAP} {\bfseries
  12} (2024) 015} [\href{https://arxiv.org/abs/2405.10155}{{\ttfamily
  2405.10155}}].

\bibitem{Sorbo:2011rz}
L.~Sorbo, \emph{{Parity violation in the Cosmic Microwave Background from a
  pseudoscalar inflaton}},
  \href{https://doi.org/10.1088/1475-7516/2011/06/003}{\emph{JCAP} {\bfseries
  06} (2011) 003} [\href{https://arxiv.org/abs/1101.1525}{{\ttfamily
  1101.1525}}].

\bibitem{Anber:2012du}
M.~M. Anber and L.~Sorbo, \emph{{Non-Gaussianities and chiral gravitational
  waves in natural steep inflation}},
  \href{https://doi.org/10.1103/PhysRevD.85.123537}{\emph{Phys. Rev. D}
  {\bfseries 85} (2012) 123537}
  [\href{https://arxiv.org/abs/1203.5849}{{\ttfamily 1203.5849}}].

\bibitem{Adshead:2013qp}
P.~Adshead, E.~Martinec and M.~Wyman, \emph{{Gauge fields and inflation: Chiral
  gravitational waves, fluctuations, and the Lyth bound}},
  \href{https://doi.org/10.1103/PhysRevD.88.021302}{\emph{Phys. Rev. D}
  {\bfseries 88} (2013) 021302}
  [\href{https://arxiv.org/abs/1301.2598}{{\ttfamily 1301.2598}}].

\bibitem{Adshead:2018doq}
P.~Adshead, J.~T. Giblin and Z.~J. Weiner, \emph{{Gravitational waves from
  gauge preheating}},
  \href{https://doi.org/10.1103/PhysRevD.98.043525}{\emph{Phys. Rev. D}
  {\bfseries 98} (2018) 043525}
  [\href{https://arxiv.org/abs/1805.04550}{{\ttfamily 1805.04550}}].

\bibitem{Adshead:2019aac}
P.~Adshead, L.~Pearce, M.~Peloso, M.~A. Roberts and L.~Sorbo,
  \emph{{Gravitational waves from fermion production during axion inflation}},
  \href{https://doi.org/10.1088/1475-7516/2019/10/018}{\emph{JCAP} {\bfseries
  10} (2019) 018} [\href{https://arxiv.org/abs/1904.10483}{{\ttfamily
  1904.10483}}].

\bibitem{Adshead:2019igv}
P.~Adshead, J.~T. Giblin, M.~Pieroni and Z.~J. Weiner, \emph{{Constraining
  Axion Inflation with Gravitational Waves across 29 Decades in Frequency}},
  \href{https://doi.org/10.1103/PhysRevLett.124.171301}{\emph{Phys. Rev. Lett.}
  {\bfseries 124} (2020) 171301}
  [\href{https://arxiv.org/abs/1909.12843}{{\ttfamily 1909.12843}}].

\bibitem{Adshead:2019lbr}
P.~Adshead, J.~T. Giblin, M.~Pieroni and Z.~J. Weiner, \emph{{Constraining
  axion inflation with gravitational waves from preheating}},
  \href{https://doi.org/10.1103/PhysRevD.101.083534}{\emph{Phys. Rev. D}
  {\bfseries 101} (2020) 083534}
  [\href{https://arxiv.org/abs/1909.12842}{{\ttfamily 1909.12842}}].

\bibitem{Bastero-Gil:2022fme}
M.~Bastero-Gil and A.~T. Manso, \emph{{Parity violating gravitational waves at
  the end of inflation}},
  \href{https://doi.org/10.1088/1475-7516/2023/08/001}{\emph{JCAP} {\bfseries
  08} (2023) 001} [\href{https://arxiv.org/abs/2209.15572}{{\ttfamily
  2209.15572}}].

\bibitem{Linde2013}
A.~Linde, S.~Mooij and E.~Pajer, \emph{Gauge field production in supergravity
  inflation: Local non-gaussianity and primordial black holes},
  \href{https://doi.org/10.1103/PhysRevD.87.103506}{\emph{Phys. Rev. D}
  {\bfseries 87} (2013) 103506}
  [\href{https://arxiv.org/abs/1212.1693}{{\ttfamily 1212.1693}}].

\bibitem{Bugaev:2013fya}
E.~Bugaev and P.~Klimai, \emph{{Axion inflation with gauge field production and
  primordial black holes}},
  \href{https://doi.org/10.1103/PhysRevD.90.103501}{\emph{Phys. Rev. D}
  {\bfseries 90} (2014) 103501}
  [\href{https://arxiv.org/abs/1312.7435}{{\ttfamily 1312.7435}}].

\bibitem{Domcke:2017fix}
V.~Domcke, F.~Muia, M.~Pieroni and L.~T. Witkowski, \emph{{PBH dark matter from
  axion inflation}},
  \href{https://doi.org/10.1088/1475-7516/2017/07/048}{\emph{JCAP} {\bfseries
  07} (2017) 048} [\href{https://arxiv.org/abs/1704.03464}{{\ttfamily
  1704.03464}}].

\bibitem{Domcke:2019mnd}
V.~Domcke, B.~von Harling, E.~Morgante and K.~Mukaida, \emph{{Baryogenesis from
  axion inflation}},
  \href{https://doi.org/10.1088/1475-7516/2019/10/032}{\emph{JCAP} {\bfseries
  10} (2019) 032} [\href{https://arxiv.org/abs/1905.13318}{{\ttfamily
  1905.13318}}].

\bibitem{Domcke:2020kcp}
V.~Domcke, Y.~Ema, K.~Mukaida and M.~Yamada, \emph{{Spontaneous Baryogenesis
  from Axions with Generic Couplings}},
  \href{https://doi.org/10.1007/JHEP08(2020)096}{\emph{JHEP} {\bfseries 08}
  (2020) 096} [\href{https://arxiv.org/abs/2006.03148}{{\ttfamily
  2006.03148}}].

\bibitem{Garretson:1992vt}
W.~D. Garretson, G.~B. Field and S.~M. Carroll, \emph{{Primordial magnetic
  fields from pseudoGoldstone bosons}},
  \href{https://doi.org/10.1103/PhysRevD.46.5346}{\emph{Phys. Rev. D}
  {\bfseries 46} (1992) 5346}
  [\href{https://arxiv.org/abs/hep-ph/9209238}{{\ttfamily hep-ph/9209238}}].

\bibitem{Anber:2006xt}
M.~M. Anber and L.~Sorbo, \emph{{N-flationary magnetic fields}},
  \href{https://doi.org/10.1088/1475-7516/2006/10/018}{\emph{JCAP} {\bfseries
  10} (2006) 018} [\href{https://arxiv.org/abs/astro-ph/0606534}{{\ttfamily
  astro-ph/0606534}}].

\bibitem{Fujita:2015iga}
T.~Fujita, R.~Namba, Y.~Tada, N.~Takeda and H.~Tashiro, \emph{{Consistent
  generation of magnetic fields in axion inflation models}},
  \href{https://doi.org/10.1088/1475-7516/2015/05/054}{\emph{JCAP} {\bfseries
  05} (2015) 054} [\href{https://arxiv.org/abs/1503.05802}{{\ttfamily
  1503.05802}}].

\bibitem{Adshead:2016iae}
P.~Adshead, J.~T. Giblin, T.~R. Scully and E.~I. Sfakianakis,
  \emph{{Magnetogenesis from axion inflation}},
  \href{https://doi.org/10.1088/1475-7516/2016/10/039}{\emph{JCAP} {\bfseries
  10} (2016) 039} [\href{https://arxiv.org/abs/1606.08474}{{\ttfamily
  1606.08474}}].

\bibitem{Kamarpour:2018ckk}
M.~Kamarpour and O.~Sobol, \emph{{Magnetogenesis in Natural Inflation Model}},
  \href{https://doi.org/10.15407/ujpe63.8.673}{\emph{Ukr. J. Phys.} {\bfseries
  63} (2018) 673}.

\bibitem{Sobol:2019xls}
O.~O. Sobol, E.~V. Gorbar and S.~I. Vilchinskii, \emph{{Backreaction of
  electromagnetic fields and the Schwinger effect in pseudoscalar inflation
  magnetogenesis}},
  \href{https://doi.org/10.1103/PhysRevD.100.063523}{\emph{Phys. Rev. D}
  {\bfseries 100} (2019) 063523}
  [\href{https://arxiv.org/abs/1907.10443}{{\ttfamily 1907.10443}}].

\bibitem{Gorbar:2021zlr}
E.~V. Gorbar, K.~Schmitz, O.~O. Sobol and S.~I. Vilchinskii,
  \emph{{Hypermagnetogenesis from axion inflation: Model-independent
  estimates}}, \href{https://doi.org/10.1103/PhysRevD.105.043530}{\emph{Phys.
  Rev. D} {\bfseries 105} (2022) 043530}
  [\href{https://arxiv.org/abs/2111.04712}{{\ttfamily 2111.04712}}].

\bibitem{Gorbar:2021rlt}
E.~V. Gorbar, K.~Schmitz, O.~O. Sobol and S.~I. Vilchinskii, \emph{{Gauge-field
  production during axion inflation in the gradient expansion formalism}},
  \href{https://doi.org/10.1103/PhysRevD.104.123504}{\emph{Phys. Rev. D}
  {\bfseries 104} (2021) 123504}
  [\href{https://arxiv.org/abs/2109.01651}{{\ttfamily 2109.01651}}].

\bibitem{Maleknejad:2012fw}
A.~Maleknejad, M.~M. Sheikh-Jabbari and J.~Soda, \emph{{Gauge Fields and
  Inflation}}, \href{https://doi.org/10.1016/j.physrep.2013.03.003}{\emph{Phys.
  Rept.} {\bfseries 528} (2013) 161}
  [\href{https://arxiv.org/abs/1212.2921}{{\ttfamily 1212.2921}}].

\bibitem{Adshead:2013nka}
P.~Adshead, E.~Martinec and M.~Wyman, \emph{{Perturbations in Chromo-Natural
  Inflation}}, \href{https://doi.org/10.1007/JHEP09(2013)087}{\emph{JHEP}
  {\bfseries 09} (2013) 087} [\href{https://arxiv.org/abs/1305.2930}{{\ttfamily
  1305.2930}}].

\bibitem{Dimastrogiovanni:2012ew}
E.~Dimastrogiovanni and M.~Peloso, \emph{{Stability analysis of chromo-natural
  inflation and possible evasion of Lyth\textquoteright{}s bound}},
  \href{https://doi.org/10.1103/PhysRevD.87.103501}{\emph{Phys. Rev. D}
  {\bfseries 87} (2013) 103501}
  [\href{https://arxiv.org/abs/1212.5184}{{\ttfamily 1212.5184}}].

\bibitem{Maleknejad:2016qjz}
A.~Maleknejad, \emph{{Axion Inflation with an SU(2) Gauge Field: Detectable
  Chiral Gravity Waves}},
  \href{https://doi.org/10.1007/JHEP07(2016)104}{\emph{JHEP} {\bfseries 07}
  (2016) 104} [\href{https://arxiv.org/abs/1604.03327}{{\ttfamily
  1604.03327}}].

\bibitem{Maleknejad:2016dci}
A.~Maleknejad, \emph{{Gravitational leptogenesis in axion inflation with SU(2)
  gauge field}},
  \href{https://doi.org/10.1088/1475-7516/2016/12/027}{\emph{JCAP} {\bfseries
  12} (2016) 027} [\href{https://arxiv.org/abs/1604.06520}{{\ttfamily
  1604.06520}}].

\bibitem{Dimastrogiovanni:2016fuu}
E.~Dimastrogiovanni, M.~Fasiello and T.~Fujita, \emph{{Primordial Gravitational
  Waves from Axion-Gauge Fields Dynamics}},
  \href{https://doi.org/10.1088/1475-7516/2017/01/019}{\emph{JCAP} {\bfseries
  01} (2017) 019} [\href{https://arxiv.org/abs/1608.04216}{{\ttfamily
  1608.04216}}].

\bibitem{Domcke:2018rvv}
V.~Domcke, B.~Mares, F.~Muia and M.~Pieroni, \emph{{Emerging chromo-natural
  inflation}}, \href{https://doi.org/10.1088/1475-7516/2019/04/034}{\emph{JCAP}
  {\bfseries 04} (2019) 034}
  [\href{https://arxiv.org/abs/1807.03358}{{\ttfamily 1807.03358}}].

\bibitem{Maleknejad:2018nxz}
A.~Maleknejad and E.~Komatsu, \emph{{Production and Backreaction of Spin-2
  Particles of $SU(2)$ Gauge Field during Inflation}},
  \href{https://doi.org/10.1007/JHEP05(2019)174}{\emph{JHEP} {\bfseries 05}
  (2019) 174} [\href{https://arxiv.org/abs/1808.09076}{{\ttfamily
  1808.09076}}].

\bibitem{Lozanov:2018kpk}
K.~D. Lozanov, A.~Maleknejad and E.~Komatsu, \emph{{Schwinger Effect by an
  $SU(2)$ Gauge Field during Inflation}},
  \href{https://doi.org/10.1007/JHEP02(2019)041}{\emph{JHEP} {\bfseries 02}
  (2019) 041} [\href{https://arxiv.org/abs/1805.09318}{{\ttfamily
  1805.09318}}].

\bibitem{Dimastrogiovanni:2018xnn}
E.~Dimastrogiovanni, M.~Fasiello, R.~J. Hardwick, H.~Assadullahi, K.~Koyama and
  D.~Wands, \emph{{Non-Gaussianity from Axion-Gauge Fields Interactions during
  Inflation}}, \href{https://doi.org/10.1088/1475-7516/2018/11/029}{\emph{JCAP}
  {\bfseries 11} (2018) 029}
  [\href{https://arxiv.org/abs/1806.05474}{{\ttfamily 1806.05474}}].

\bibitem{Wolfson:2020fqz}
I.~Wolfson, A.~Maleknejad and E.~Komatsu, \emph{{How attractive is the
  isotropic attractor solution of axion-SU(2) inflation?}},
  \href{https://doi.org/10.1088/1475-7516/2020/09/047}{\emph{JCAP} {\bfseries
  09} (2020) 047} [\href{https://arxiv.org/abs/2003.01617}{{\ttfamily
  2003.01617}}].

\bibitem{Iarygina:2021bxq}
O.~Iarygina and E.~I. Sfakianakis, \emph{{Gravitational waves from spectator
  Gauge-flation}},
  \href{https://doi.org/10.1088/1475-7516/2021/11/023}{\emph{JCAP} {\bfseries
  11} (2021) 023} [\href{https://arxiv.org/abs/2105.06972}{{\ttfamily
  2105.06972}}].

\bibitem{Fujita:2022jkc}
T.~Fujita, K.~Imagawa and K.~Murai, \emph{{Gravitational waves detectable in
  laser interferometers from axion-SU(2) inflation}},
  \href{https://doi.org/10.1088/1475-7516/2022/07/046}{\emph{JCAP} {\bfseries
  07} (2022) 046} [\href{https://arxiv.org/abs/2203.15273}{{\ttfamily
  2203.15273}}].

\bibitem{Iarygina:2023mtj}
O.~Iarygina, E.~I. Sfakianakis, R.~Sharma and A.~Brandenburg,
  \emph{{Backreaction of axion-SU(2) dynamics during inflation}},
  \href{https://doi.org/10.1088/1475-7516/2024/04/018}{\emph{JCAP} {\bfseries
  04} (2024) 018} [\href{https://arxiv.org/abs/2311.07557}{{\ttfamily
  2311.07557}}].

\bibitem{Fujita:2023axo}
T.~Fujita, K.~Murai, K.~Nakayama and W.~Yin, \emph{{Misalignment production of
  vector boson dark matter from axion-SU(2) inflation}},
  \href{https://doi.org/10.1088/1475-7516/2024/04/007}{\emph{JCAP} {\bfseries
  04} (2024) 007} [\href{https://arxiv.org/abs/2312.06889}{{\ttfamily
  2312.06889}}].

\bibitem{Dimastrogiovanni:2024xvc}
E.~Dimastrogiovanni, M.~Fasiello and A.~Papageorgiou, \emph{{Novel primordial
  black hole production mechanism from non-Abelian gauge fields during
  inflation}}, \href{https://doi.org/10.1103/PhysRevD.110.103542}{\emph{Phys.
  Rev. D} {\bfseries 110} (2024) 103542}
  [\href{https://arxiv.org/abs/2403.13581}{{\ttfamily 2403.13581}}].

\bibitem{Dimastrogiovanni:2024lzj}
E.~Dimastrogiovanni, M.~Fasiello, M.~Michelotti and O.~\"Ozsoy, \emph{{A
  universal constraint on axion non-Abelian dynamics during inflation}},
  \href{https://doi.org/10.1088/1475-7516/2025/03/007}{\emph{JCAP} {\bfseries
  03} (2025) 007} [\href{https://arxiv.org/abs/2405.17411}{{\ttfamily
  2405.17411}}].

\bibitem{Badger:2024ekb}
C.~Badger, H.~Duval, T.~Fujita, S.~Kuroyanagi, A.~Romero-Rodr'iguez and
  M.~Sakellariadou, \emph{{Detection prospects of gravitational waves from
  SU(2) axion inflation}},
  \href{https://doi.org/10.1103/PhysRevD.110.084063}{\emph{Phys. Rev. D}
  {\bfseries 110} (2024) 084063}
  [\href{https://arxiv.org/abs/2406.11742}{{\ttfamily 2406.11742}}].

\bibitem{Cheng:2015oqa}
S.-L. Cheng, W.~Lee and K.-W. Ng, \emph{{Numerical study of pseudoscalar
  inflation with an axion-gauge field coupling}},
  \href{https://doi.org/10.1103/PhysRevD.93.063510}{\emph{Phys. Rev. D}
  {\bfseries 93} (2016) 063510}
  [\href{https://arxiv.org/abs/1508.00251}{{\ttfamily 1508.00251}}].

\bibitem{Domcke:2020zez}
V.~Domcke, V.~Guidetti, Y.~Welling and A.~Westphal, \emph{{Resonant
  backreaction in axion inflation}},
  \href{https://doi.org/10.1088/1475-7516/2020/09/009}{\emph{JCAP} {\bfseries
  09} (2020) 009} [\href{https://arxiv.org/abs/2002.02952}{{\ttfamily
  2002.02952}}].

\bibitem{Peloso:2022ovc}
M.~Peloso and L.~Sorbo, \emph{{Instability in axion inflation with strong
  backreaction from gauge modes}},
  \href{https://doi.org/10.1088/1475-7516/2023/01/038}{\emph{JCAP} {\bfseries
  01} (2023) 038} [\href{https://arxiv.org/abs/2209.08131}{{\ttfamily
  2209.08131}}].

\bibitem{Sobol:2020lec}
O.~O. Sobol, A.~V. Lysenko, E.~V. Gorbar and S.~I. Vilchinskii, \emph{{Gradient
  expansion formalism for magnetogenesis in the kinetic coupling model}},
  \href{https://doi.org/10.1103/PhysRevD.102.123512}{\emph{Phys. Rev. D}
  {\bfseries 102} (2020) 123512}
  [\href{https://arxiv.org/abs/2010.13587}{{\ttfamily 2010.13587}}].

\bibitem{Caravano:2022epk}
A.~Caravano, E.~Komatsu, K.~D. Lozanov and J.~Weller, \emph{{Lattice
  simulations of axion-U(1) inflation}},
  \href{https://doi.org/10.1103/PhysRevD.108.043504}{\emph{Phys. Rev. D}
  {\bfseries 108} (2023) 043504}
  [\href{https://arxiv.org/abs/2204.12874}{{\ttfamily 2204.12874}}].

\bibitem{Figueroa:2023oxc}
D.~G. Figueroa, J.~Lizarraga, A.~Urio and J.~Urrestilla, \emph{{Strong
  Backreaction Regime in Axion Inflation}},
  \href{https://doi.org/10.1103/PhysRevLett.131.151003}{\emph{Phys. Rev. Lett.}
  {\bfseries 131} (2023) 151003}
  [\href{https://arxiv.org/abs/2303.17436}{{\ttfamily 2303.17436}}].

\bibitem{Caravano:2024xsb}
A.~Caravano and M.~Peloso, \emph{{Unveiling the nonlinear dynamics of a rolling
  axion during inflation}},  \href{https://arxiv.org/abs/2407.13405}{{\ttfamily
  2407.13405}}.

\bibitem{Domcke:2023tnn}
V.~Domcke, Y.~Ema and S.~Sandner, \emph{{Perturbatively including
  inhomogeneities in axion inflation}},
  \href{https://doi.org/10.1088/1475-7516/2024/03/019}{\emph{JCAP} {\bfseries
  03} (2024) 019} [\href{https://arxiv.org/abs/2310.09186}{{\ttfamily
  2310.09186}}].

\bibitem{Khlopov:2008qy}
M.~Y. Khlopov, \emph{{Primordial Black Holes}},
  \href{https://doi.org/10.1088/1674-4527/10/6/001}{\emph{Res. Astron.
  Astrophys.} {\bfseries 10} (2010) 495}
  [\href{https://arxiv.org/abs/0801.0116}{{\ttfamily 0801.0116}}].

\bibitem{Sasaki:2018dmp}
M.~Sasaki, T.~Suyama, T.~Tanaka and S.~Yokoyama, \emph{{Primordial black
  holes\textemdash{}perspectives in gravitational wave astronomy}},
  \href{https://doi.org/10.1088/1361-6382/aaa7b4}{\emph{Class. Quant. Grav.}
  {\bfseries 35} (2018) 063001}
  [\href{https://arxiv.org/abs/1801.05235}{{\ttfamily 1801.05235}}].

\bibitem{Escriva:2022duf}
A.~Escriv\`a, F.~Kuhnel and Y.~Tada, \emph{{Primordial Black Holes}},
  \href{https://arxiv.org/abs/2211.05767}{{\ttfamily 2211.05767}}.

\bibitem{Mishra:2019pzq}
S.~S. Mishra and V.~Sahni, \emph{{Primordial Black Holes from a tiny bump/dip
  in the Inflaton potential}},
  \href{https://doi.org/10.1088/1475-7516/2020/04/007}{\emph{JCAP} {\bfseries
  04} (2020) 007} [\href{https://arxiv.org/abs/1911.00057}{{\ttfamily
  1911.00057}}].

\bibitem{Ragavendra:2020sop}
H.~V. Ragavendra, P.~Saha, L.~Sriramkumar and J.~Silk, \emph{{Primordial black
  holes and secondary gravitational waves from ultraslow roll and punctuated
  inflation}}, \href{https://doi.org/10.1103/PhysRevD.103.083510}{\emph{Phys.
  Rev. D} {\bfseries 103} (2021) 083510}
  [\href{https://arxiv.org/abs/2008.12202}{{\ttfamily 2008.12202}}].

\bibitem{Domcke:2018eki}
V.~Domcke and K.~Mukaida, \emph{{Gauge Field and Fermion Production during
  Axion Inflation}},
  \href{https://doi.org/10.1088/1475-7516/2018/11/020}{\emph{JCAP} {\bfseries
  11} (2018) 020} [\href{https://arxiv.org/abs/1806.08769}{{\ttfamily
  1806.08769}}].

\bibitem{Leblond:2010yq}
L.~Leblond and E.~Pajer, \emph{{Resonant Trispectrum and a Dozen More
  Primordial N-point functions}},
  \href{https://doi.org/10.1088/1475-7516/2011/01/035}{\emph{JCAP} {\bfseries
  01} (2011) 035} [\href{https://arxiv.org/abs/1010.4565}{{\ttfamily
  1010.4565}}].

\bibitem{Caravano:2022yyv}
A.~Caravano, \emph{{Simulating the inflationary Universe: from single-field to
  the axion-U(1) model}}, Ph.D. thesis, Munich U., Munich U., 7, 2022.
\newblock \href{https://arxiv.org/abs/2209.13616}{{\ttfamily 2209.13616}}.
\newblock 10.5282/edoc.30905.

\bibitem{Adshead:2023mvt}
P.~Adshead, J.~T. Giblin, R.~Grutkoski and Z.~J. Weiner, \emph{{Gauge
  preheating with full general relativity}},
  \href{https://doi.org/10.1088/1475-7516/2024/03/017}{\emph{JCAP} {\bfseries
  03} (2024) 017} [\href{https://arxiv.org/abs/2311.01504}{{\ttfamily
  2311.01504}}].

\bibitem{Galanti:2024jhw}
D.~C. Galanti, P.~Conzinu, G.~Marozzi and S.~Santos~da Costa, \emph{{Gauge
  invariant quantum backreaction in U(1) axion inflation}},
  \href{https://arxiv.org/abs/2406.19960}{{\ttfamily 2406.19960}}.

\bibitem{JOSS}
{\scshape Pencil Code} Collaboration, A.~{Brandenburg}, A.~{Johansen},
  P.~{Bourdin}, W.~{Dobler}, W.~{Lyra}, M.~{Rheinhardt}, S.~{Bingert},
  N.~{Haugen}, A.~{Mee}, F.~{Gent}, N.~{Babkovskaia}, C.-C. {Yang},
  T.~{Heinemann}, B.~{Dintrans}, D.~{Mitra}, S.~{Candelaresi}, J.~{Warnecke},
  P.~{K{\"a}pyl{\"a}}, A.~{Schreiber}, P.~{Chatterjee}, M.~{K{\"a}pyl{\"a}},
  X.-Y. {Li}, J.~{Kr{\"u}ger}, J.~{Aarnes}, G.~{Sarson}, J.~{Oishi},
  J.~{Schober}, R.~{Plasson}, C.~{Sandin}, E.~{Karchniwy}, L.~{Rodrigues},
  A.~{Hubbard}, G.~{Guerrero}, A.~{Snodin}, I.~{Losada}, J.~{Pekkil{\"a}} and
  C.~{Qian}, \emph{{The Pencil Code, a modular MPI code for partial
  differential equations and particles: multipurpose and
  multiuser-maintained}}, \href{https://doi.org/10.21105/joss.02807}{\emph{J.
  Open Source Software} {\bfseries 6} (2021) 2807}.

\bibitem{Figueroa:2024rkr}
D.~G. Figueroa, J.~Lizarraga, N.~Loayza, A.~Urio and J.~Urrestilla, \emph{{The
  non-linear dynamics of axion inflation: a detailed lattice study}},
  \href{https://arxiv.org/abs/2411.16368}{{\ttfamily 2411.16368}}.

\bibitem{Pagano:2015hma}
L.~Pagano, L.~Salvati and A.~Melchiorri, \emph{{New constraints on primordial
  gravitational waves from Planck 2015}},
  \href{https://doi.org/10.1016/j.physletb.2016.07.078}{\emph{Phys. Lett. B}
  {\bfseries 760} (2016) 823}
  [\href{https://arxiv.org/abs/1508.02393}{{\ttfamily 1508.02393}}].

\bibitem{Abazajian:2019eic}
K.~Abazajian et~al., \emph{{CMB-S4 Science Case, Reference Design, and Project
  Plan}},  \href{https://arxiv.org/abs/1907.04473}{{\ttfamily 1907.04473}}.

\bibitem{Neronov:1900zz}
A.~Neronov and I.~Vovk, \emph{{Evidence for strong extragalactic magnetic
  fields from Fermi observations of TeV blazars}},
  \href{https://doi.org/10.1126/science.1184192}{\emph{Science} {\bfseries 328}
  (2010) 73} [\href{https://arxiv.org/abs/1006.3504}{{\ttfamily 1006.3504}}].

\bibitem{Taylor:2011bn}
A.~M. Taylor, I.~Vovk and A.~Neronov, \emph{{Extragalactic magnetic fields
  constraints from simultaneous GeV-TeV observations of blazars}},
  \href{https://doi.org/10.1051/0004-6361/201116441}{\emph{Astron. Astrophys.}
  {\bfseries 529} (2011) A144}
  [\href{https://arxiv.org/abs/1101.0932}{{\ttfamily 1101.0932}}].

\bibitem{Finke:2013tyq}
{\scshape Fermi-LAT} Collaboration, J.~Finke, L.~Reyes and M.~Georganopoulos,
  \emph{{Constraints on the Intergalactic Magnetic Field from Gamma-Ray
  Observations of Blazars}}, {\emph{eConf} {\bfseries C121028} (2012) 365}
  [\href{https://arxiv.org/abs/1303.5093}{{\ttfamily 1303.5093}}].

\bibitem{Finke:2015ona}
J.~D. Finke, L.~C. Reyes, M.~Georganopoulos, K.~Reynolds, M.~Ajello, S.~J.
  Fegan and K.~McCann, \emph{{Constraints on the Intergalactic Magnetic Field
  with Gamma-Ray Observations of Blazars}},
  \href{https://doi.org/10.1088/0004-637X/814/1/20}{\emph{Astrophys. J.}
  {\bfseries 814} (2015) 20}
  [\href{https://arxiv.org/abs/1510.02485}{{\ttfamily 1510.02485}}].

\bibitem{Turner:1987bw}
M.~S. Turner and L.~M. Widrow, \emph{{Inflation Produced, Large Scale Magnetic
  Fields}}, \href{https://doi.org/10.1103/PhysRevD.37.2743}{\emph{Phys. Rev. D}
  {\bfseries 37} (1988) 2743}.

\bibitem{Ratra:1991bn}
B.~Ratra, \emph{{Cosmological `seed' magnetic field from inflation}},
  \href{https://doi.org/10.1086/186384}{\emph{Astrophys. J. Lett.} {\bfseries
  391} (1992) L1}.

\bibitem{Demozzi:2009fu}
V.~Demozzi, V.~Mukhanov and H.~Rubinstein, \emph{{Magnetic fields from
  inflation?}},
  \href{https://doi.org/10.1088/1475-7516/2009/08/025}{\emph{JCAP} {\bfseries
  08} (2009) 025} [\href{https://arxiv.org/abs/0907.1030}{{\ttfamily
  0907.1030}}].

\bibitem{Martin:2007ue}
J.~Martin and J.~Yokoyama, \emph{{Generation of Large-Scale Magnetic Fields in
  Single-Field Inflation}},
  \href{https://doi.org/10.1088/1475-7516/2008/01/025}{\emph{JCAP} {\bfseries
  01} (2008) 025} [\href{https://arxiv.org/abs/0711.4307}{{\ttfamily
  0711.4307}}].

\bibitem{Ferreira:2013sqa}
R.~J.~Z. Ferreira, R.~K. Jain and M.~S. Sloth, \emph{{Inflationary
  magnetogenesis without the strong coupling problem}},
  \href{https://doi.org/10.1088/1475-7516/2013/10/004}{\emph{JCAP} {\bfseries
  10} (2013) 004} [\href{https://arxiv.org/abs/1305.7151}{{\ttfamily
  1305.7151}}].

\bibitem{Sharma:2017eps}
R.~Sharma, S.~Jagannathan, T.~R. Seshadri and K.~Subramanian, \emph{{Challenges
  in Inflationary Magnetogenesis: Constraints from Strong Coupling,
  Backreaction and the Schwinger Effect}},
  \href{https://doi.org/10.1103/PhysRevD.96.083511}{\emph{Phys. Rev. D}
  {\bfseries 96} (2017) 083511}
  [\href{https://arxiv.org/abs/1708.08119}{{\ttfamily 1708.08119}}].

\bibitem{Kushwaha:2020nfa}
A.~Kushwaha and S.~Shankaranarayanan, \emph{{Helical magnetic fields from
  Riemann coupling}},
  \href{https://doi.org/10.1103/PhysRevD.102.103528}{\emph{Phys. Rev. D}
  {\bfseries 102} (2020) 103528}
  [\href{https://arxiv.org/abs/2008.10825}{{\ttfamily 2008.10825}}].

\bibitem{Kushwaha:2022bwy}
A.~Kushwaha, A.~Naskar, D.~Nandi and S.~Shankaranarayanan, \emph{{Effective
  field theory of magnetogenesis identify necessary and sufficient
  conditions}},
  \href{https://doi.org/10.1088/1475-7516/2023/01/045}{\emph{JCAP} {\bfseries
  01} (2023) 045} [\href{https://arxiv.org/abs/2207.05162}{{\ttfamily
  2207.05162}}].

\bibitem{Maity:2021qps}
D.~Maity, S.~Pal and T.~Paul, \emph{{Effective Theory of Inflationary
  Magnetogenesis and Constraints on Reheating}},
  \href{https://doi.org/10.1088/1475-7516/2021/05/045}{\emph{JCAP} {\bfseries
  05} (2021) 045} [\href{https://arxiv.org/abs/2103.02411}{{\ttfamily
  2103.02411}}].

\bibitem{Tripathy:2021sfb}
S.~Tripathy, D.~Chowdhury, R.~K. Jain and L.~Sriramkumar, \emph{{Challenges in
  the choice of the nonconformal coupling function in inflationary
  magnetogenesis}},
  \href{https://doi.org/10.1103/PhysRevD.105.063519}{\emph{Phys. Rev. D}
  {\bfseries 105} (2022) 063519}
  [\href{https://arxiv.org/abs/2111.01478}{{\ttfamily 2111.01478}}].

\bibitem{Brandenburg:2024awd}
A.~Brandenburg, O.~Iarygina, E.~I. Sfakianakis and R.~Sharma,
  \emph{{Magnetogenesis from axion-SU(2) inflation}},
  \href{https://arxiv.org/abs/2408.17413}{{\ttfamily 2408.17413}}.

\bibitem{Hogan1983}
C.~J. Hogan, \emph{Magnetohydrodynamic effects of a first-order cosmological
  phase transition},
  \href{https://doi.org/10.1103/PhysRevLett.51.1488}{\emph{Phys. Rev. Lett.}
  {\bfseries 51} (1983) 1488}.

\bibitem{Quashnock1989}
J.~M. {Quashnock}, A.~{Loeb} and D.~N. {Spergel}, \emph{{Magnetic Field
  Generation during the Cosmological QCD Phase Transition}},
  \href{https://doi.org/10.1086/185528}{\emph{The Astrophysical Journal
  Letters} {\bfseries 344} (1989) L49}.

\bibitem{Vachaspati:1991nm}
T.~Vachaspati, \emph{{Magnetic fields from cosmological phase transitions}},
  \href{https://doi.org/10.1016/0370-2693(91)90051-Q}{\emph{Phys. Lett. B}
  {\bfseries 265} (1991) 258}.

\bibitem{Baym:1995fk}
G.~Baym, D.~Bodeker and L.~D. McLerran, \emph{{Magnetic fields produced by
  phase transition bubbles in the electroweak phase transition}},
  \href{https://doi.org/10.1103/PhysRevD.53.662}{\emph{Phys. Rev. D} {\bfseries
  53} (1996) 662} [\href{https://arxiv.org/abs/hep-ph/9507429}{{\ttfamily
  hep-ph/9507429}}].

\bibitem{Sigl:1996dm}
G.~Sigl, A.~V. Olinto and K.~Jedamzik, \emph{{Primordial magnetic fields from
  cosmological first order phase transitions}},
  \href{https://doi.org/10.1103/PhysRevD.55.4582}{\emph{Phys. Rev. D}
  {\bfseries 55} (1997) 4582}
  [\href{https://arxiv.org/abs/astro-ph/9610201}{{\ttfamily
  astro-ph/9610201}}].

\bibitem{Durrer:2013pga}
R.~Durrer and A.~Neronov, \emph{{Cosmological Magnetic Fields: Their
  Generation, Evolution and Observation}},
  \href{https://doi.org/10.1007/s00159-013-0062-7}{\emph{Astron. Astrophys.
  Rev.} {\bfseries 21} (2013) 62}
  [\href{https://arxiv.org/abs/1303.7121}{{\ttfamily 1303.7121}}].

\bibitem{Subramanian:2015lua}
K.~Subramanian, \emph{{The origin, evolution and signatures of primordial
  magnetic fields}},
  \href{https://doi.org/10.1088/0034-4885/79/7/076901}{\emph{Rept. Prog. Phys.}
  {\bfseries 79} (2016) 076901}
  [\href{https://arxiv.org/abs/1504.02311}{{\ttfamily 1504.02311}}].

\bibitem{Vachaspati:2020blt}
T.~Vachaspati, \emph{{Progress on cosmological magnetic fields}},
  \href{https://doi.org/10.1088/1361-6633/ac03a9}{\emph{Rept. Prog. Phys.}
  {\bfseries 84} (2021) 074901}
  [\href{https://arxiv.org/abs/2010.10525}{{\ttfamily 2010.10525}}].

\bibitem{Bran+Prot23}
A.~{Brandenburg} and N.~N. {Protiti}, \emph{{Electromagnetic Conversion into
  Kinetic and Thermal Energies}},
  \href{https://doi.org/10.3390/e25091270}{\emph{Entropy} {\bfseries 25} (2023)
  1270} [\href{https://arxiv.org/abs/2308.00662}{{\ttfamily 2308.00662}}].

\bibitem{BEO96}
A.~{Brandenburg}, K.~{Enqvist} and P.~{Olesen}, \emph{{Large-scale magnetic
  fields from hydromagnetic turbulence in the very early universe}},
  \href{https://doi.org/10.1103/PhysRevD.54.1291}{\emph{Phys. Rev. D}
  {\bfseries 54} (1996) 1291}
  [\href{https://arxiv.org/abs/astro-ph/9602031}{{\ttfamily
  astro-ph/9602031}}].

\bibitem{Mattias2001}
M.~Christensson, M.~Hindmarsh and A.~Brandenburg, \emph{Inverse cascade in
  decaying three-dimensional magnetohydrodynamic turbulence},
  \href{https://doi.org/10.1103/PhysRevE.64.056405}{\emph{Phys. Rev. E}
  {\bfseries 64} (2001) 056405}.

\bibitem{Banerjee:2004df}
R.~Banerjee and K.~Jedamzik, \emph{{The Evolution of cosmic magnetic fields:
  From the very early universe, to recombination, to the present}},
  \href{https://doi.org/10.1103/PhysRevD.70.123003}{\emph{Phys. Rev. D}
  {\bfseries 70} (2004) 123003}
  [\href{https://arxiv.org/abs/astro-ph/0410032}{{\ttfamily
  astro-ph/0410032}}].

\bibitem{Brandenburg:2016odr}
A.~Brandenburg and T.~Kahniashvili, \emph{{Classes of hydrodynamic and
  magnetohydrodynamic turbulent decay}},
  \href{https://doi.org/10.1103/PhysRevLett.118.055102}{\emph{Phys. Rev. Lett.}
  {\bfseries 118} (2017) 055102}
  [\href{https://arxiv.org/abs/1607.01360}{{\ttfamily 1607.01360}}].

\bibitem{Adshead:2015pva}
P.~Adshead, J.~T. Giblin, T.~R. Scully and E.~I. Sfakianakis,
  \emph{{Gauge-preheating and the end of axion inflation}},
  \href{https://doi.org/10.1088/1475-7516/2015/12/034}{\emph{JCAP} {\bfseries
  12} (2015) 034} [\href{https://arxiv.org/abs/1502.06506}{{\ttfamily
  1502.06506}}].

\bibitem{MAGIC:2022piy}
{\scshape MAGIC} Collaboration, V.~A. Acciari et~al., \emph{{A lower bound on
  intergalactic magnetic fields from time variability of 1ES 0229+200 from
  MAGIC and Fermi/LAT observations}},
  \href{https://doi.org/10.1051/0004-6361/202244126}{\emph{Astron. Astrophys.}
  {\bfseries 670} (2023) A145}
  [\href{https://arxiv.org/abs/2210.03321}{{\ttfamily 2210.03321}}].

\bibitem{Zhou2019}
M.~{Zhou}, P.~{Bhat}, N.~F. {Loureiro} and D.~A. {Uzdensky}, \emph{{Magnetic
  island merger as a mechanism for inverse magnetic energy transfer}},
  \href{https://doi.org/10.1103/PhysRevResearch.1.012004}{\emph{Phys. Rev.
  Res.} {\bfseries 1} (2019) 012004}
  [\href{https://arxiv.org/abs/1901.02448}{{\ttfamily 1901.02448}}].

\bibitem{Bhat2021}
P.~{Bhat}, M.~{Zhou} and N.~F. {Loureiro}, \emph{{Inverse energy transfer in
  decaying, three-dimensional, non-helical magnetic turbulence due to magnetic
  reconnection}}, \href{https://doi.org/10.1093/mnras/staa3849}{\emph{Mon. Not.
  Roy. Astron. Soc.} {\bfseries 501} (2021) 3074}
  [\href{https://arxiv.org/abs/2007.07325}{{\ttfamily 2007.07325}}].

\bibitem{Hosking:2020wom}
D.~N. Hosking and A.~A. Schekochihin, \emph{{Reconnection-Controlled Decay of
  Magnetohydrodynamic Turbulence and the Role of Invariants}},
  \href{https://doi.org/10.1103/PhysRevX.11.041005}{\emph{Phys. Rev. X}
  {\bfseries 11} (2021) 041005}
  [\href{https://arxiv.org/abs/2012.01393}{{\ttfamily 2012.01393}}].

\bibitem{Hosking:2022umv}
D.~N. Hosking and A.~A. Schekochihin, \emph{{Cosmic-void observations
  reconciled with primordial magnetogenesis}},
  \href{https://doi.org/10.1038/s41467-023-43258-3}{\emph{Nature Commun.}
  {\bfseries 14} (2023) 7523}
  [\href{https://arxiv.org/abs/2203.03573}{{\ttfamily 2203.03573}}].

\bibitem{Brandenburg:2024tyi}
A.~Brandenburg, A.~Neronov and F.~Vazza, \emph{{Resistively controlled
  primordial magnetic turbulence decay}},
  \href{https://doi.org/10.1051/0004-6361/202449267}{\emph{Astron. Astrophys.}
  {\bfseries 687} (2024) A186}
  [\href{https://arxiv.org/abs/2401.08569}{{\ttfamily 2401.08569}}].

\bibitem{Schekochihin+22}
A.~A. {Schekochihin}, \emph{{MHD turbulence: a biased review}},
  \href{https://doi.org/10.1017/S0022377822000721}{\emph{J. Plasma Phys.}
  {\bfseries 88} (2022) 155880501}
  [\href{https://arxiv.org/abs/2010.00699}{{\ttfamily 2010.00699}}].

\bibitem{Kernan:1995bz}
P.~J. Kernan, G.~D. Starkman and T.~Vachaspati, \emph{{Big bang nucleosynthesis
  constraints on primordial magnetic fields}},
  \href{https://doi.org/10.1103/PhysRevD.54.7207}{\emph{Phys. Rev. D}
  {\bfseries 54} (1996) 7207}
  [\href{https://arxiv.org/abs/astro-ph/9509126}{{\ttfamily
  astro-ph/9509126}}].

\bibitem{2016A&A...594A..19P}
P.~Collaboration, \emph{{Planck 2015 results. XIX. Constraints on primordial
  magnetic fields}},
  \href{https://doi.org/10.1051/0004-6361/201525821}{\emph{Astron. Astrophys.}
  {\bfseries 594} (2016) A19}
  [\href{https://arxiv.org/abs/1502.01594}{{\ttfamily 1502.01594}}].

\bibitem{Trivedi:2013wqa}
P.~Trivedi, K.~Subramanian and T.~R. Seshadri, \emph{{Primordial magnetic field
  limits from the CMB trispectrum: Scalar modes and Planck constraints}},
  \href{https://doi.org/10.1103/PhysRevD.89.043523}{\emph{Phys. Rev. D}
  {\bfseries 89} (2014) 043523}
  [\href{https://arxiv.org/abs/1312.5308}{{\ttfamily 1312.5308}}].

\bibitem{1974ApJ...187..425P}
W.~H. {Press} and P.~{Schechter}, \emph{{Formation of Galaxies and Clusters of
  Galaxies by Self-Similar Gravitational Condensation}},
  \href{https://doi.org/10.1086/152650}{\emph{The Astrophysical Journal}
  {\bfseries 187} (1974) 425}.

\bibitem{Anne2004}
A.~M. Green, A.~R. Liddle, K.~A. Malik and M.~Sasaki, \emph{New calculation of
  the mass fraction of primordial black holes},
  \href{https://doi.org/10.1103/PhysRevD.70.041502}{\emph{Phys. Rev. D}
  {\bfseries 70} (2004) 041502}.

\bibitem{Byrnes:2012yx}
C.~T. Byrnes, E.~J. Copeland and A.~M. Green, \emph{{Primordial black holes as
  a tool for constraining non-Gaussianity}},
  \href{https://doi.org/10.1103/PhysRevD.86.043512}{\emph{Phys. Rev. D}
  {\bfseries 86} (2012) 043512}
  [\href{https://arxiv.org/abs/1206.4188}{{\ttfamily 1206.4188}}].

\bibitem{PhysRev.82.664}
J.~Schwinger, \emph{On gauge invariance and vacuum polarization},
  \href{https://doi.org/10.1103/PhysRev.82.664}{\emph{Phys. Rev.} {\bfseries
  82} (1951) 664}.

\bibitem{Frob:2014zka}
M.~B. Fr\"ob, J.~Garriga, S.~Kanno, M.~Sasaki, J.~Soda, T.~Tanaka and
  A.~Vilenkin, \emph{{Schwinger effect in de Sitter space}},
  \href{https://doi.org/10.1088/1475-7516/2014/04/009}{\emph{JCAP} {\bfseries
  04} (2014) 009} [\href{https://arxiv.org/abs/1401.4137}{{\ttfamily
  1401.4137}}].

\bibitem{Kobayashi:2014zza}
T.~Kobayashi and N.~Afshordi, \emph{{Schwinger Effect in 4D de Sitter Space and
  Constraints on Magnetogenesis in the Early Universe}},
  \href{https://doi.org/10.1007/JHEP10(2014)166}{\emph{JHEP} {\bfseries 10}
  (2014) 166} [\href{https://arxiv.org/abs/1408.4141}{{\ttfamily 1408.4141}}].

\bibitem{sharma2017}
R.~Sharma and S.~Singh, \emph{{Multifaceted Schwinger effect in de Sitter
  space}}, \href{https://doi.org/10.1103/PhysRevD.96.025012}{\emph{Phys. Rev.
  D} {\bfseries 96} (2017) 025012}
  [\href{https://arxiv.org/abs/1704.05076}{{\ttfamily 1704.05076}}].

\bibitem{Fujita:2022fwc}
T.~Fujita, J.~Kume, K.~Mukaida and Y.~Tada, \emph{{Effective treatment of U(1)
  gauge field and charged particles in axion inflation}},
  \href{https://doi.org/10.1088/1475-7516/2022/09/023}{\emph{JCAP} {\bfseries
  09} (2022) 023} [\href{https://arxiv.org/abs/2204.01180}{{\ttfamily
  2204.01180}}].

\bibitem{vonEckardstein:2024tix}
R.~von Eckardstein, K.~Schmitz and O.~Sobol, \emph{{On the Schwinger effect
  during axion inflation}},  \href{https://arxiv.org/abs/2408.16538}{{\ttfamily
  2408.16538}}.

\bibitem{DATA}
\emph{{Datasets of ``Lattice simulations of axion-U(1) inflation: gravitational
  waves, magnetic fields, and scalar statistics",
  \href{https://zenodo.org/records/15074254}{doi: 10.5281/zenodo.15074254}
  (v2025.03.24)}; see also
  \url{http://norlx65.nordita.org/~brandenb/projects/axion-U1/} for easier
  access}, .

\bibitem{RoperPol+20}
A.~{Roper Pol}, A.~{Brandenburg}, T.~{Kahniashvili}, A.~{Kosowsky} and
  S.~{Mandal}, \emph{{The timestep constraint in solving the gravitational wave
  equations sourced by hydromagnetic turbulence}},
  \href{https://doi.org/10.1080/03091929.2019.1653460}{\emph{Geophys.
  Astrophys. Fluid Dyn.} {\bfseries 114} (2020) 130}
  [\href{https://arxiv.org/abs/1807.05479}{{\ttfamily 1807.05479}}].

\end{thebibliography}\endgroup
\end{document}